\documentclass[aps,amsmath,preprint,showpacs,showkeys]{revtex4}
\usepackage{graphicx}
\usepackage{amsmath}
\usepackage{epigraph}
\usepackage[T1]{fontenc}
\usepackage{pslatex}
\usepackage{color}
\begin{document}
\def\b{\textcolor{black}}
\def\la{{\langle}}
\def\ra{{\rangle}}
\def\vep{{\varepsilon}}
\newcommand{\beq}{\begin{equation}}
\newcommand{\eeq}{\end{equation}}
\newcommand{\beqa}{\begin{eqnarray}}
\newcommand{\eeqa}{\end{eqnarray}}
\newcommand{\q}{\quad}
\newcommand{\tunn}{\text{tunn}}
\newcommand{\refl}{\text{refl}}
\newcommand{\all}{\text{all}}
\newcommand{\ion}{\text{ion}}
\newcommand{\bound}{\text{bound}}
\newcommand{\free}{\text{free}}
\newcommand{\ur}{\uparrow}
\newcommand{\dr}{\downarrow}
\newcommand{\lc}{\curly{l}}
\newcommand{\C}{\hat{C}}
\newcommand{\upp}{\hat{\underline{p}}}
\newcommand{\B}{\hat{B}}
\newcommand{\HH}{\hat{H}}
\newcommand{\pb}{\hat{\pi}^B}
\newcommand{\A}{\tilde{A}}
\newcommand{\Aa}{\hat{A}}
\newcommand{\pc}{\hat{\pi}^C}
\newcommand{\pd}{\hat{\pi}^D}
\newcommand{\D}{\hat{D}}
\newcommand{\Pii}{\hat{\Pi}}
\newcommand{\s}{\hat{S}}
\newcommand{\x}{x_{cl}}
\newcommand{\si}{\hat{\sigma}}
\newcommand{\Pp}{\hat{\Pi}}
\newcommand{\AC}{{\it AC }}
\newcommand{\pha}{{\phi_A}}
\newcommand{\phb}{{\phi_B}}
\newcommand{\La}{{\lambda }}
\newcommand{\psii}{|\psi_I\ra}
\newcommand{\psif}{|\psi_F\ra}
\newcommand{\n}{\\ \nonumber}
\newcommand{\nn}{\q\q\q\q\q\q\q\q\q\\ \nonumber}
\q\q\q\q\q\q\q\q\q
\newcommand{\om}{\omega}
\newcommand{\U}{\hat{U}}
\newcommand{\up}{\hat{U}_{part}}
\newcommand{\mf}{m_f^{\alpha}}
\newcommand{\e}{\epsilon}
\newcommand{\Om}{\Omega}
\newcommand{\Tau}{\mathcal{T}_{SWP}}
\newcommand{\Ttu}{\tau_{in/out}}
\newcommand{\br}{\overline}
\newcommand{\cn}[1]{#1_{\hbox{\scriptsize{con}}}}
\newcommand{\sy}[1]{#1_{\hbox{\scriptsize{sys}}}}
\newcommand{\PAD }{Pad\'{e}\q}
\newcommand{\PP }{\hat{\Pi}}
\newcommand{\su}{{\phi \gets \psi}}
\newcommand{\get }{\leftarrow}
\newcommand{\Eq}{equation }
\newcommand{\f}{\ref }
\newcommand{\T}{\text{T}_\Om}
\newcommand{\Tf}{\text{T}}
\newcommand{{\ttau}}{\overline{\tau_\Om} }
\newcommand{{\tttu}}{\overline{\tau_{[0,d]}} }
\newcommand{\h}{\hat{H}}
\newcommand{\N}{\mathfrak{N} }
\newcommand{\I}{\text{Im } }

\title{Path probabilities for consecutive measurements, and certain "quantum paradoxes"}
\author {D. Sokolovski$^{1,2}$}
\affiliation {$^1$ Departmento de Qu\'imica-F\'isica, Universidad del Pa\' is Vasco, UPV/EHU, Leioa, Spain}
\affiliation{$^2$ IKERBASQUE, Basque Foundation for Science, Maria Diaz de Haro 3, 48013, Bilbao, Spain}
\date{\today}
\begin{abstract}
\noindent
ABSTRACT: 
\newline 
{We consider a finite-dimensional quantum system, making a transition between known initial and final states.
The outcomes of several accurate measurements, which {\it could be} made in the interim,
define virtual paths, each endowed with a probability amplitude.
If the measurements are {\it actually made}, the paths, which may now be called "real",  acquire also the probabilities, 
related to the frequencies, with which a  path is seen to be travelled in a series of identical trials. 
Different sets of measurements, made on the same system, can produce 
different, or incompatible, statistical ensembles, whose conflicting attributes 
may, although by no means should, appear "paradoxical". 
We describe in detail the ensembles, resulting from intermediate measurements of mutually commuting, or non-commuting, operators, 
in terms of the real paths produced. In the same manner, we analyse the Hardy's and the "three box"  paradoxes, 
the photon's past in an interferometer, the "quantum Cheshire cat" experiment, as well as the closely related subject of 
"interaction-free measurements".  It is shown that, in all these cases, inaccurate "weak measurements" produce no real paths, 
and yield only limited information about the virtual paths' probability amplitudes.}  
\end{abstract}
\keywords{Quantum measurements, Feynman paths, quantum "paradoxes"}
\maketitle
\vskip0.5cm
\section{Introduction}
Recently, there has been significant interest in the properties of a pre-and post-selected quantum systems, 
and, in particular, in the description of such systems during the time between the preparation, and
the arrival in the pre-determined final state (see, for example \cite{Chin} and the Refs. therein). Intermediate state of the system can be probed by 
performing, one after another, measurements of various physical quantities.
Although produced from the same quantum system, statistical ensembles, resulting from different sets of measurements, are known to have 
conflicting and seemingly incompatible qualities. These conflicts have, in turn, led to the discussion of certain "quantum paradoxes",
allegedly specific to a system, subjected to post-selection. 
Such is, for example, the "three box paradox" \cite{3Ba}-\cite{3Bd}, claiming that a particle
can be, at the same time, at two different locations "with certainty". A similarly "paradoxical" suggestion that a photon could, on its way to detection,
have visited the places it had "never entered, nor left, was made in \cite{Nest1}-\cite{Nest2}, and further discussed in  \cite{Nest3}-\cite{Nest6}. In the discussion of the Hardy's paradox \cite{HARDY1}-\cite{HARDY4} the particle 
is suspected of simultaneously "being and not being" at the same location \cite{HARDY2}. The so called "quantum Cheshire cat" scheme \cite{CAT1}-\cite{CAT3}
promises "disembodiment of physical properties from the object they belong to". 
\newline
One can easily dismiss a "paradox"  of this type simply by noting that the conflicting features are never observed in the same experimental setup, and therefore, never occur "simultaneously" \cite{3Bd}, \cite{HARDY4}, \cite{Stapp1}, \cite{Unruh}, \cite{Stapp2}, \cite{Kastner} .
(We agree: one can use a piece of plasticine to make a ball, or a cube, but should not claim that an object can be a ball and a cube at the same time.) There have been attempts to ascertain "simultaneous presence" of the conflicting attributes by subjecting the system 
to weakly perturbing, or "weak" measurements \cite{3Ba}, \cite{Nest1}, \cite{HARDY2}, \cite{CAT1}. However, such measurements only probe the values of 
the relative probability amplitudes, corresponding to the processes of interest \cite{CAT4}, \cite{CAT5}, \cite{DSann} and by no means prove that these processes 
are, indeed, taking place at the same time. Furthermore, confusing these amplitudes with the value of a physical quantity, 
may lead to such unhelpful concepts as "negative numbers of particles" \cite{HARDY2}, "negative durations" spent by a non-relativistic particle in a specified region of space \cite{TRAV}. or "apparently superluminal"
transmission of a tunnelling particle across a potential barrier \cite{Ann}.
\newline
\b{Similar questions can be asked about macroscopic quantum systems, such as superconducting flux qubits \cite{LG1}. 
The answers are often formulated in terms of the Leggett-Garg inequalities \cite{LG2}, which restrict the values of  correlators of physical observables, under the assumption that macroscopic superpositions cannot persist for some fundamental 
reason (for a review see \cite{LG3}, and for recent developments \cite{LG4}).}
\newline
It is not our intention to compile an exhaustive list of relevant  literature, or to discuss all aspects of the subject in great detail.
The main purpose of this paper is 
to describe consecutive quantum measurements in a simple language, relying only on the most basic principles and concepts of elementary quantum mechanics. The brief introduction, already made, may have helped to convince the reader that  such a description would indeed be desirable.
\newline
We start from a simple premise. With the initial and final states of the system fixed, there are many measurements which could, 
in principle, be performed in the interim. Connecting results of  possible measurements,
performed at different times,
defines a {\it virtual} path, which a system could follow. For each virtual path 
quantum mechanics provides a complex valued probability amplitude $A(path)$.
If the said measurements are actually made, the outcomes become a sequence of observable events, 
the path acquires a probability, $P(path)=|A(path)|^2$,
and becomes {\it real}. (We use "real" as a natural complement to "virtual".) 
The set of all real paths, possible final destinations, and the corresponding probabilities,  together define a classical statistical ensemble. 
We note that a similar ensemble could, in principle, be constructed also by purely classical means. 
For example, it is not difficult to imagine a gun, whose bullets would arrive at a point on the screen 
with exactly the same probability, as the electrons in the Young's double slit experiment. 
For someone, interested {\it only} in the number of particles per square centimetre, 
the two ensembles would be identical.
\newline
The quantum side of the discussion is, therefore, limited to the details of the ensemble's preparation.
In the previous example, the same distribution can be achieved by aiming the gun at different angles, 
and varying the frequency of the shots, or by using quantum interference. 
In the following, we will be interested in exploiting the quantum properties of a system, and will {return} 
to a classical analogue only occasionally. 
\newline
Two well known features of conventional quantum measurements \cite{Real3},
 are easily translated into the language of quantum paths.
The fact that the results of a measurement cannot "pre-exist" the measurement  \cite{Merm}, means that different sets of measurements 
are able to produce different, and in some sense incompatible, statistical ensembles.
The fact  that quantum mechanics is non-local was summarised in \cite{Merm} as follows:
{\it "...the value assigned to an observable must depend on the complete experimental arrangement
under which it is measured, even when two arrangements differ only far from the region in which the value is ascertained.}
Thus, we can expect that a single change, made in a distant part of the setup, may produce a new network of real paths, all parts of which will be incompatible 
 with the original ensemble. 
  \newline
Throughout the rest of the paper, we will attempt to make the proposed description of consecutive 
 quantum measurements sufficiently precise, explore its usefulness, and review, with its help, some of the quantum "paradoxes",
 already mentioned above.
 The rest of the paper is organised as follows. 
 In Sect. II we continue with the introduction, and revisit Feynman's example \cite{Feyns}, used to demonstrate  that quantum probabilities cannot 
 pre-exist the measurement, and must be produced  in its course.
 In Sect. III  we describe one way to decide whether two given sets of averages can belong to the same statistical ensemble.
 In Sect. IV we show that the results in Sect. I fail the test of Sect. II, and would require the use of at least two different ensembles, 
in order to be reproduced by purely classical means.  
In Sect. V we show how a statistical ensemble, 
describing a system which can reach its final destination in several different ways,  is produced by applying a sequence of accurate von Neumann measurements.
 In Sect. VI we consider an example, in which the real paths are produced by measurements of commuting operators, acting in a three-dimensional Hilbert space.
 In Section VII we introduce non-perturbing (non-demolition) measurements, capable of providing a more detailed description of a real path, without altering the ensemble. 
 In Sect. VIII we briefly discuss the "three-box" paradox. 
 In Section IX we return to the question of photon's past in an interferometer containing a nested loop.
 In Sect. X we review the related case of "interaction-free measurements",  
 discuss its similarity to the double-slit experiment, 
 and describe the minimal requirements for a classical setup, designed to reproduce the quantum behaviour. 
In Section XI we extend our analysis to composite systems, and analyse the Hardy's paradox in terms of quantum paths.
Section XII contains a minimalist version of "quantum Cheshire cat" experiment, analysed it terms 
of the real paths produced by the measurements.
In Sect. XIII we demonstrate that the "weak measurements" only yield limited information about 
probability amplitudes and, therefore, provide little insight beyond what is already known.
Section XIV contains our conclusions.
\section{Feynman's example.}
Feynman \cite{Feyns} proposed the following example in order to demonstrate the impossibility of imitating quantum physics by purely classical means. Alice  (A) and Bob (B) are given each a spin-1/2 particle of a pair produced in an entangled state 
 \begin{eqnarray}\label{a1}
|\Phi\ra =\left [|\ur^A_z\ra|\ur^B_z\ra+|\dr_z^A\ra|\dr^B_z\ra\right ]/\sqrt{2},
\end{eqnarray}
by a remote source operated by Carol (C).
(Here $|\ur(\dr)^{A,B}_z\ra$ are the spin states of Alice's or Bob's  particle, polarised up (down) the $z$-axis.)
Both make accurate spin measurements along the axis making angles $\pha$ (Alice) and $\phb$ (Bob) with the $z$-axis in the $xz$-plane. There are four possible outcomes, with each spin pointing either up ($|\ur_\phi\ra$) or down ($|\dr_\phi\ra$) their respective axis.
An elementary quantum calculation shows that the corresponding joint probabilities  are given by
 \begin{eqnarray}\label{a2}
W(\ur_\pha\ur_\phb)=W(\dr_\pha\dr_\phb)=\cos^2(\pha-\phb)/2, \n
W(\ur_\pha\dr_\phb)=W(\ur_\pha\dr_\phb)=\sin^2(\pha-\phb)/2.
\end{eqnarray}
The point of the exercise is to demonstrate that the statistics in Eq.(\ref{a2}) are inconsistent with the assumption that particles arrive to Alice and Bob in pre-determine spin states, which then uniquely determine the outcome of each individual measurement. 
To do so, it is sufficient 
to consider only seven possible angles $\pha=0, \pi/6, \pi/3,\pi/2,2\pi/3,5\pi/6$, which Alice will choose randomly, with equal probabilities. With Bob always measuring at $\phb=\pha+\pi/6$, the probability to have the same result is
 \begin{eqnarray}\label{a2a}
W(same)=W(\ur_\pha\ur_\phb)+W(\dr_\pha\dr_\phb)=3/4.
\end{eqnarray}
To prove the point, we assume that, in the classical version of the experiment,  Carol sends Alice and Bob the same instruction, containing a $7$-digit strings of $0$'s and $1$'s. {With the results of all possible measurements now pre-determined,
Alice randomly chooses the $j$-th position in her string, $j=1,2,..,6$, which corresponds to measuring her spin at $\pha=(j-1)\pi/6$.
Then she and Bob read the $j$-th and $(j+1)$-th entries, respectively, and count the number of cases, $N(same)$, in which their
results coincide. After $N>>1$ trials, the ratio $N(same)/N$ gives a good approximation to the classical probability $W'(same)$.
The question is whether Carol can achieve $W'(same)=3/4$, as would be the case in the quantum version of the game.  
\newline
Not all instructions are, however, admissible. In the quantum case, Bob measuring at $\pha+\pi/2$, 
will always disagree with Alice, since from Eq.(\ref{a2}) we have $W(\ur_\pha\ur_\phb)=W(\dr_\pha\dr_\phb)=0$.
To mimic this in the classical case, we must ensure that the entries at the $j$-th and $j+3$-d positions disagree as well.
This leaves Carol with only eight possible strings, which she can send with the probabilities }
$p_i\ge 0$, $i=1,2,...,8$, $\sum_{i=1}^8 p_i=1$, 
 \begin{eqnarray}\label{a3}
\text{Prob}[1110001]=p_1,\q \text{Prob}[0001110]=p_5,\n
\text{Prob}[1100011]=p_2,\q \text{Prob}[0011100]=p_6,\n
\text{Prob}[1010101]=p_3,\q \text{Prob}[0101010]=p_7,\n
\text{Prob}[0111000]=p_4,\q \text{Prob}[1000111]=p_8.\n
\end{eqnarray}
{Equations (\ref{a3}) define a simple statistical ensemble with eight possible states (instructions), each endowed with the corresponding probability. Carol's task is now to find a suitable set of $p_i$'s, such that A and B will have the same result with a
 probability given by Eq.(\ref{a2a}).}
\newline
This, however turns out to be impossible. The best rate of coincidences 
 is achieved if Carol sends only the instruction(s) with the largest possible number of agreements, $N_{max}$, between a digit and its neighbour to the right. Such are the states $1$ and $5$ in Eqs.(\ref{a3}),
for which we have $N_{max}=4$. With Alice choosing her entry randomly, the classical probability for her and Bob's  results to coincide is limited by 
 \begin{eqnarray}\label{a3a}
W^{class}(same)\le N_{max}/6=2/3.
\end{eqnarray}
Disagreement with the correct quantum result (\ref{a2a}) is evident,  so the classical simulation must fail. 
\section{Compatibility test. Negative probability.}
We are not interested in merely repeating the argument of \cite{Feyns}, and will go one step further, to look at the kind of difficulties which would occur, should we try to reconcile the  probabilities $p_j$ in (\ref{a3}) with those in Eq.(\ref{a2}). 
Consider first  $N$ states $x_i$, $i=1,..,N$  and a corresponding set of probabilities, $p_i \ge 0$, 
There is also a set of known real functions $F^k(x_i)=(F^k_1, F^k_2, ...., F^k_n)$, $k=1,...,K$.
Alice, in her new role, evaluates the averages
 \begin{eqnarray}\label{s1}
\la F^k\ra \equiv \sum_{i=1}^N F^k_ip_i,\q k=1,...,K
\end{eqnarray}
 by sampling  an $x_i$ according to the distribution $p_i$, writing down $F(x_i)$, adding up all the results, 
 and dividing the sum by the number of trials. In doing so, Alice may use the same distribution $p_i$ 
or, perhaps, choose a different one, $p'_i$ for some of the $\la F^k\ra$. Given the values of $\la F^k\ra$, we wish to know whether Alice 
 could have used the same distribution for all her averages, or had to apply  a different one at least once.
 \newline 
 To answer this question we need to examine $K+1$ linear equations,
 \begin{eqnarray}\label{s2}
 \sum_{i=1}^N p_i = 1, \q\q\q\q\q\q\q\q \n
\sum_{i=1}^N F^k_ip_i = \la F^k\ra,\q k=1,...,K,
\end{eqnarray}
 in order to see if  they admit nonnegative solutions $p_i \ge 0$. 
 If they do, Alice could have used the $p_i$ for all $F^k$. If not, some of the averages $\la F^k\ra$ must 
 correspond to a different statistical ensemble. We will call results $\la F^k\ra$ {\it incompatible}, 
 if one or more different ensembles are required for their preparation.
 \newline A trivial example is the case $N=K=2$, $F^1=(1,0)$, $F^2=(0,1)$. If the averages, suppled by Alice,
 are $\la F^1\ra = 0.4$ and $\la F^2\ra = 0.7$, equations (\ref{s2}) have no solutions.
Thus, we 
 conclude that Alice must have used $p_1=1-p_2=0.4$, when averaging $F^1$, and a different 
 $p'_2=1-p'_1=0.7$ for averaging $F^2$.
 \newline
Another example involves  $N=3$, $K=2$, $F^1=(1,0,0)$, and $F^2=(0,0,1)$.
If Alice supplies the values $\la F^1\ra=1$ and $\la F^2\ra=1$, Eqs.(\ref{s2}) have a unique, yet unsuitable, solution
 \begin{eqnarray}\label{s3}
p_1=p_3=-p_2=1,  
\end{eqnarray}
With $p_2=-1$, Alice would not be able to realise the ensemble (\ref{s3}) in practice \cite{Feyns}.
Thus, we must conclude that she had to use the set $p_i=\delta_{i1}$ when experimenting with $F^1$, and a different set $p'_i=\delta_{i3}$,
when averaging $F^2$. In the following we will use this simple test in order to establish whether different 
results of quantum measurements may or may not correspond to the same classical statistical ensemble. 
\section{Different ensembles and non-locality}
Next we return to the example of Sect.II, and examine it in more detail.
The probabilities $W^{(\pha,\phb)}(same)$
 of Alice and Bob obtaining the same results when measuring at $(\pha=0,\pi/6,\pi/3)$ and $ (\phb=\pi/6,\pi/3,\pi/2)$, 
 respectively, can be written as the averages, 
  \begin{eqnarray}\label{t1}
W^{(\pha,\phb)}(same)=\sum_{i=1}^8 F_i^{(\pha,\phb)}p_i, 
\end{eqnarray}
where
  \begin{eqnarray}\label{t2}
F^{(0,\pi/6)}=(11001100),\q
F^{(\pi/6,\pi/3)}=(10011001),\q
\text{and} \q
F^{(\pi/3,\pi/2)}=(01010101). 
\end{eqnarray}
Equating $W^{(\pha,\phb) }(same)$ to the quantum result $W(same)=W(\ur_\pha\ur_\phb)+W(\dr_\pha\dr_\phb)=3/4$ yields  Eqs. (\ref{s2}) in the form
  \begin{eqnarray}\label{t3}
p_1+p_2+p_3+p_4+p_5+p_6+p_7+p_8=1,\n
p_1+p_2+p_5+p_6=3/4,\q\q\q\q\q\q\q\n
p_1+p_4+p_5+p_8=3/4,\q\q\q\q\q\q\q\n
p_2+p_4+p_6+p_8=3/4.\q\q\q\q\q\q\q 
\end{eqnarray}
It is readily seen that the quantum results at $(\pha=0, \phb=\pi/6)$ and $(\pha=\pi/6, \phb=\pi/3)$ are compatible, and can be simulated classically, e.g., by choosing
\begin{eqnarray}\label{t4}
 p_1=p_5=3/8, \q p_2=p_4=p_6=p_8=0,\q 
 p_3=p_7=1/8,\q\q\q\q\q\q\q\q\q\q,
\end{eqnarray}
as a valid solution of the first three equations in (\ref{t3}).
It is, however, impossible to find a set 
$0\le p_i\le 1$, $i=1,..,8$,  which would satisfy all four equations in (\ref{t3}). Indeed, 
adding the last three equations, and subtracting the result  from the first one, we have  $p_3+p_7=-1/8 <0$.
\newline
We {\it can}, however, reproduce quantum results classically, if we change the rules of the experiment.
 To do so, Bob would only need to inform Carol
that a measurement at $(\pha=\pi/3, \phb=\pi/2)$ is to be performed next, and the $p_i$'s in (\ref{t4})  that were suitable 
for $(\pha=0, \phb=\pi/6)$ and $(\pha=\pi/6, \phb=\pi/3)$
must be changed to some $p'_i$s, which would give the correct results for $(\pha=\pi/3, \phb=\pi/2)$. In other words, Carol would need to be told when to use a different statistical ensemble (\ref{a3}).
The peculiarity of the quantum case is in the fact that Alice and Bob are able achieve the same result 
locally, by agreeing only between themselves the angles for their respective measurements. There is no need for a direct communication with the Carol, since that task appears to be taken care of by the non-local quantum correlations implied by the entangled state (\ref{a1}).
\begin{figure} 
	\centering
		\includegraphics[width=7cm,height=5cm]{{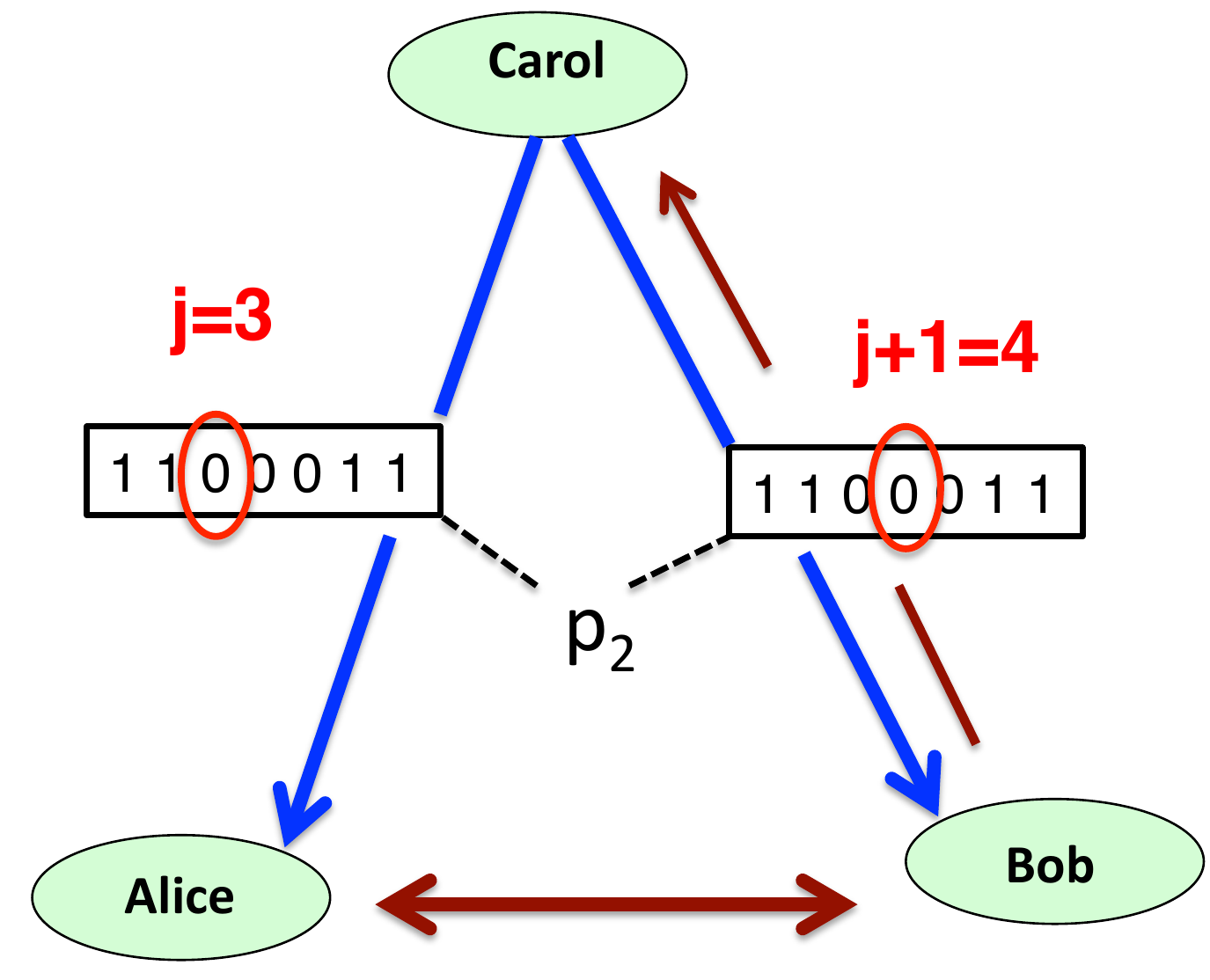}}
\caption{(Colour online) A classical scheme, whereby Carol (C) chooses the $i$-th instruction from the set (\ref{a3})  with a probability 
$p_i$, and sends it to both Alice (A)  and Bob (B). First, A chooses at random an integer $1\le j \le 6$,
and communicates it to B. Then A and B read the $j$-th and $(j+1)$-th digits of their instruction, respectively, 
and check if they coincide. It is not possible to reproduce the rate of coincidence predicted for the quantum version of the experiment, 
unless A's choice of $j$ is communicated to C, who then adjusts the probabilities $p_i$ accordingly.}  
\label{fig:1}
\end{figure}
\noindent
\newline
In summary, it is possible for different sets of quantum measurements to produce incompatible averages, 
By the same token, we must conclude that at least two different statistical ensembles were used in their production.
Such ensembles cannot pre-exist the measurements, and are instead fabricated (constructed, shaped) by the 
interaction with the meters, monitoring the system. 
\section{The making of an ensemble}
To see how a statistical ensemble can be produced in practice, consider a quantum system with a Hamiltonian $\HH$, in an $N$-dimensional Hilbert space, 
prepared in an initial state $\psii$ at some $t=t_0$. Suppose we want to accurately measure,
 at three different times, $t_0\le t_1\le t_2\le t_3$,
 three 
operators $\B$, $\C$ and $\D$, 
with the eigenstates $|b_i\ra$, $|c_i\ra$ and $|d_i\ra$, $i=1,2,...,N$.
Out of the $N$ eigenvalues  $B^i$, $C^i$ and $D^i$, only
$N_B$, $N_C$, $N_D \le N$ have different values, which we will distinguish  by means of a subscript, e.g., $B_s$, $s=1,2,..,N_B$. Thus, we can write $\B$ as 
 \begin{eqnarray}\label{b1}
\B=\sum_{i=1}^{N}|b_i\ra B^i\la b_i|=\sum_{s=1}^{N_B}B_s\pb_s, \q
 \pb_s\equiv \sum_{i=1}^N \Delta (B_s-B^i)|b_i\ra\la b_i|,
\end{eqnarray}
where $\Delta (x-y)=1$ for $x=y$ and $0$ otherwise, and $ \pb_s$ is a projector onto the subspace, 
corresponding to the eigenvalue $B_s$. The operators do not necessarily commute, and 
the bases $|b_i\ra$, $|c_i\ra$ and $|d_i\ra$ are, in general, different.
\newline
The measurements will be instantaneous ones, performed by three von Neumann pointers (for details see Appendix A), whose positions are $f_B$, $f_C$ and $f_D$, 
respectively. The pointers are prepared in the states $|G_k\ra$, $k=B,C,D$, such that
 \begin{eqnarray}\label{b1a}
|G_k (f_k)|^2\equiv |\la f_k|G_k\ra|^2 = \delta (f_k), \q k=B,C,D,
\end{eqnarray}
where $\delta(f)$ is the Dirac delta.
\newline
Thus, just after all three meters have fired, the state of the composite system $|\Phi(t_3+0)\ra$ is given by
 \begin{eqnarray}\label{b3}
\la f_D|\la f_C|\la f_B|\Phi\ra=\sum_{s_D=1}^{N_D}\sum_{s_C=1}^{N_C}\sum_{s_B=1}^{N_B}
G_D(f_D-D_{s_D}) \pd_{s_D} \U(t_3,t_2)\q\q\n
\times G_C(f_C-C_{s_C}) \pc_{s_C} \U(t_2,t_1)
G_B(f_B-B_{s_B}) \pb_{s_B}\U(t_1,t_0)|\psi_I\ra,\q
\end{eqnarray}
where $\U(t_2,t_1)=\exp\left[-i\int_{t_1}^{t_2} \HH(t)dt \right ]$ is the system's evolution operator.
In the end we examine the positions of all three pointers.
\newline
 Our choice of measurements defines a statistical ensemble. There are at most $N_B\times N_C\times N_D$ possible outcomes ($B_{s_B}, C_{s_C}, D_{s_D}$), occurring with the probabilities
$P(B_{s_B}, C_{s_C}, D_{s_D})$, defined by
 \begin{eqnarray}\label{b5}
\la\Phi|f_B\ra| f_C|\ra |f_D\ra\la f_D|\la f_C|\la f_B|\Phi\ra=
\sum_{s_D=1}^{N_D}\sum_{s_C=1}^{N_C}\sum_{s_B=1}^{N_B}
\delta(f_B-B_{s_B})
\delta(f_C-C_{s_C})\delta(f_D-D_{s_D})\n
\times P(B_{s_B}, C_{s_C}, D_{s_D}).
\end{eqnarray}
\newline
Our aim is to write the probabilities $P(B_{s_B}, C_{s_C}, D_{s_D})$ in terms of the corresponding probability amplitudes.
Let $A(i_D,i_C,i_B|\psi_I)$ be the { amplitude} for "reaching the state $|d_{i_D}\ra$ at $t_3$ by passing first through the state $|b_{i_B}\ra$ at $t_1$, and then through the state $|c_{i_C}\ra$ at $t_2$ ", 
 \begin{eqnarray}\label{b2}
A(i_D,i_C,i_B|\psi_I)\equiv
\la d_{i_D}|\U(t_3,t_2)|c_{i_C}\ra
\la c_{i_C}|\U(t_2,t_1)|b_{i_B}\ra\la b_{i_B}|\U(t_1,t_0)|\psi_I\ra.\q
\end{eqnarray}
From these we can construct the { amplitudes}  $A(i_D,s_C,s_B|\psi_I)$ for "reaching a state $|d_{i_D}\ra$ at $t_3$ by passing through sub-spaces $s_B$ at $t_1$ and $s_C$ at $t_2$ ",
 \begin{eqnarray}\label{b4}
\A(i_D,s_C,s_B|\psi_I)=\sum_{i_C,i_B=1}^N
 \Delta (C_{s_C}-C^{i_C})
 \Delta (B_{s_B}-B^{i_B}) A(i_D,i_C,i_B|\psi_I).
\end{eqnarray}
%
Then for 
$P(B_{s_B}, C_{s_C}, D_{s_D})$ in Eq.(\ref{b5}) we have 
 \begin{eqnarray}\label{b6}
P(B_{s_B}, C_{s_C}, D_{s_D})=
\sum_{i_D=1}^N \Delta(D^{i_D}-D_{s_D})
p(B_{s_B}, C_{s_C}, D^{i_D}),\q
\end{eqnarray}
where
 \begin{eqnarray}\label{b6a}
p(B_{s_B}, C_{s_C}, D^{i_D})=
|\tilde A(i_D,s_C,s_B|\psi_I)|^2\equiv \n
\end{eqnarray}
is the {\it probability} of "reaching a state $|d_{i_D}\ra$ at $t_3$ by passing through sub-spaces $s_B$ at $t_1$ and $s_C$ at $t_2$"
\newline
Equations (\ref{b6}) and (\ref{b6a}) are our main result so far. They show that quantum mechanics provides $N^3$ virtual pathways ($|d_{i_d}\ra \gets
|c_{i_C}\ra \gets|b_{i_B}\ra \gets|\psi_I\ra$), connecting $|\psi_I\ra$ with $|d_{i_D}\ra$. Each pathway is endowed with a probability amplitude (\ref{b2}), which may or may not be zero. To obtain the probability that in the end the three pointers will point at  $B_{s_B}$, $C_{s_C}$, $D_{s_D}$, respectively, we must

(i) add up the amplitudes (\ref{b2}) for passing through all the states corresponding to the degenerate eigenvalues 
$B_{s_B}$, and $C_{s_C}$.

(ii) add up absolute squares of the results, $p(B_{s_B}, C_{s_C}, D^{i_D})$, summing over all states $|d_{i_D}\ra$, 
which correspond to the degenerate eigenvalue $D_{s_D}$. 
\newline
Thus, the resulting classical ensemble consists of $N$ final states, each of which can be reached via at most $N_B\times N_C$ {\it real pathways}, to which we can ascribe both the probabilities  $p(B_{s_B}, C_{s_C}, D^{i_D})$, and the amplitudes $A(i_D,s_C,s_B|\psi_I)$. 
\newline
The probabilities add up to unity, 
$\sum_{i_D=1}^N\sum_{s_C=1}^{N_C}\sum_{s_B=1}^{N_B} p(B_{s_B}, C_{s_C}, D^{i_D})=1$, which is 
guaranteed by the unitarity of the evolution of the composite, consisting of the pointers and the measured system.
The net probability to have a value $B_1$, regardless of what other values might be, 
is given by $P(B_1)=\sum_{i_D=1}^N\sum_{s_C=1}^{N_C}p(B_{1}, C_{s_C}, D^{i_D})$,
and similarly for $\C$. 
\newline
The quantum origin of the ensemble is evident from the manner in which it was designed and prepared,
Firstly, the observed probabilities must be constructed from probability amplitudes, unknown to classical physics.
Secondly, simply by choosing different degeneracies $N_B$ and $N_C$ we can combine the virtual pathways into different "real" ones, and fabricate different statistical ensembles.
Thirdly, quantum mechanics treats the "present", i.e., the time just after $t_3$, and the "past", $t<t_3$, differently.
To obtain the correct probabilities for $\B$ and $\C$, we add probability amplitudes as in Eq.(\ref{b4}), and take the absolute square of the result. We cannot do the same for $\D$ by writing $P(B_{s_B}, C_{s_C}, D_{s_D})$ as
$|\sum_{i_D=1}^N \Delta(D^{i_D}-D_{s_D})
\tilde A(i_D,s_C,s_B|\psi_I)|^2$, and should add the probabilities $p(B_{s_B}, C_{s_C}, D^{i_D})$, as in Eq.(\ref{b6}), instead. 
Finally, none of the bases $|b_i\ra$, $|c_i\ra$ and $|d_i\ra$ are defined uniquely, 
unless the eigenvalues of all three operators are non-degenerate, $N_B=N_C=N_D=N$.
Thus choosing different basis vectors to span the orthogonal subspaces, corresponding to degenerate eigenvalues
of $\B$, $\C$ and $\D$, would lead to the same observed probabilities $P(B_{s_B}, C_{s_C}, D_{s_D})$ (\ref{b6}), yet to different sets of the virtual paths in Eq.(\ref{b2}). These rules are easily generalised to a larger number of intermediate measurements.
\newline
In summary, a statistical ensemble,
produced by a series of consecutive measurements, can be described in terms of the real pathways, leading to available final states.
The number of pathways can be increased, by destroying interference between the virtual paths involved, 
or decreased, by restoring interference, where it has previously been destroyed. Both tasks can be achieved 
by choosing to measure, at intermediate times, different operators, with different numbers of degenerate eigenvalues.
\section{A three-state example}
To provide a simple illustration, we consider a three-level system, $N=3$, and limit ourselves to the intermediate measurements of mutually commuting operators, which also commute with the system's Hamiltonian, $[\B,\C]=[\B,\HH]=[\C,\HH]=0$.
Now $\B$ and $\C$ share the same eigenstates, $|c_i\ra=|b_i\ra$ and the Hamiltonian does not allow for transitions between them, since we have $\la b_i|\U(t',t)|b_j\ra=\delta_{ij}\exp[-iE_j(t'-t)]$. There are, therefore, nine virtual paths shown in Fig. 2,
which we will denote by curly brackets,
 $\{...\}$.
Omitting, where possible, the subscripts $B$, $C$, and $D$, we write
 \begin{eqnarray}\label{c1}
\{m,n\}:\q|d_m\ra \gets  |b_n\ra\gets |b_n\ra \gets |\psi_I\ra,\q m,n=1,2,3,\q\q
\end{eqnarray}
with the corresponding amplitudes given by  [cf. Eq.(\ref{b2})] 
 \begin{eqnarray}\label{c2}
A(m,n,|\psi_I)=\la d_{m}|b_{n}\ra\la b_n|\psi_I\ra
\exp[-iE_n(t_3-t_0)].
\end{eqnarray}
Such a system mimics, for example,  the situation of a particle (photon), whose wave packet is split three ways 
between the wave guides (optical fibres). 
\begin{figure}
	\centering
		\includegraphics[width=9cm,height=6cm]{{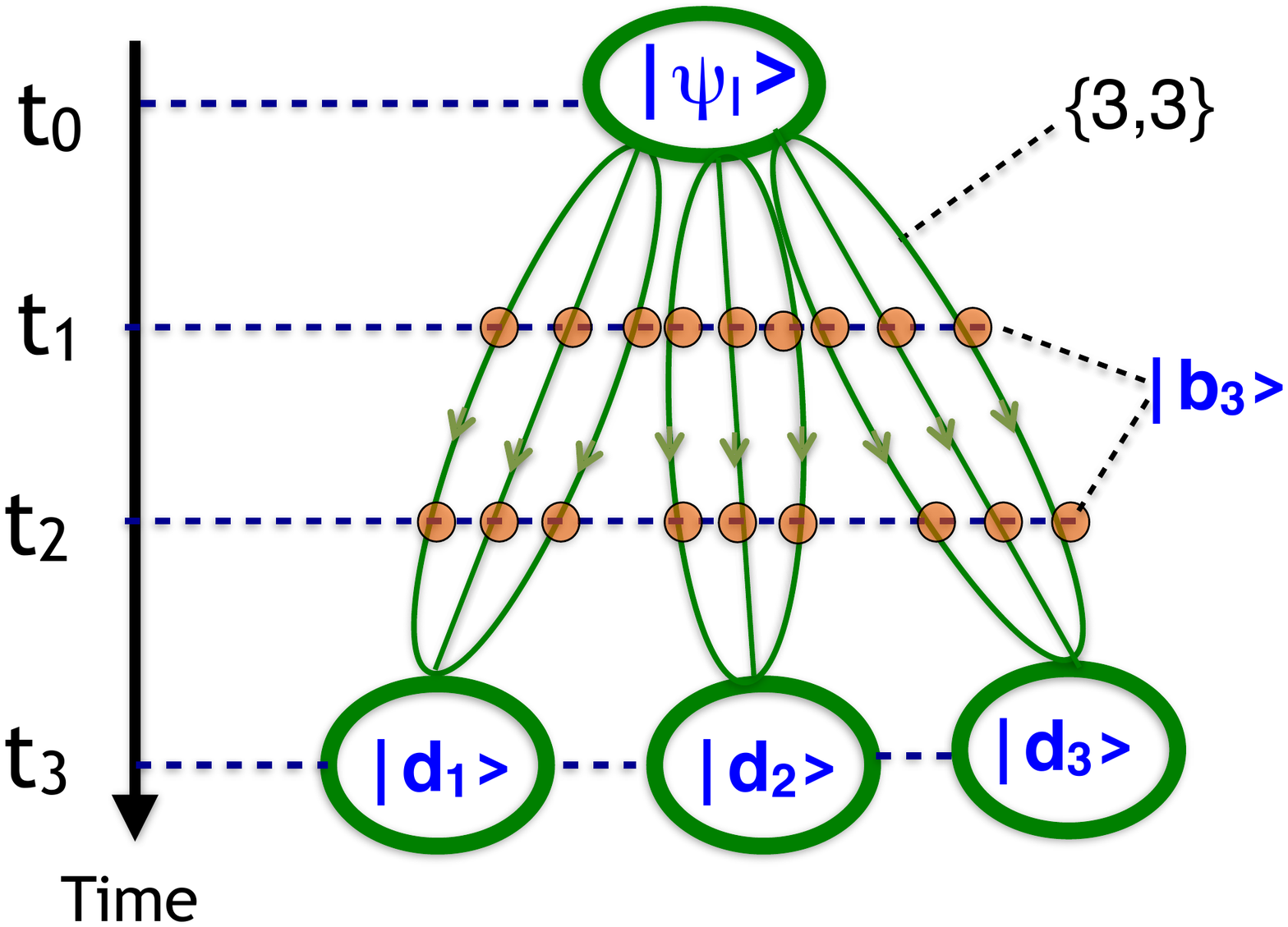}}
\caption{(Colour online) A three-level system is able reach each of the three final destinations $|d_m\ra$, $m=1,2,3$,
via three virtual paths
$\{m,n\}: 
|d_m\ra \gets |b_n\ra
\gets |b_n\ra
 \gets |\Psi_I\ra$, 
$n=1,2,3$. The number of real pathways, produced by measurements of of commuting operators $\hat B$ and $\hat C$ at $t_1$ and $t_2$, depends on the multiplicity of the operator's eigenvalues, as shown in Fig. 3. }
\label{fig:2}
\end{figure}
\newline
Suppose first that, at $t=t_1$, we accurately measure 
an operator $\B$ with two identical eigenvalues, so that its 
matrix in the $|b_i\ra$ basis is $\B=\text{diag}(B^1,B^2,B^2)$. Then at $t=t_3$ we measure an operator $\D$, which does not commute with $\B$,  and has non-degenerate eigenvalues $D^1\ne D^2\ne D^3$.
This alone creates six real paths leading to three different destinations $|d_m\ra$, namely $\{m,1\}$ and the superpositions (unions)  $\{m,2 \cup m,3\}$, so that the probabilities of the outcomes are given by
 \begin{eqnarray}\label{c3}
p(B^1,D^m)\equiv |A(m,1,|\psi_I)|^2, \q m=1,2,3,\q\q\q\n
 p(B^2,D^m)\equiv |A(m,2,|\psi_I)+A(m,3,|\psi_I)|^2.\q\q\q
\end{eqnarray}
\begin{figure}
	\centering
		\includegraphics[width=8cm,height=6cm]{{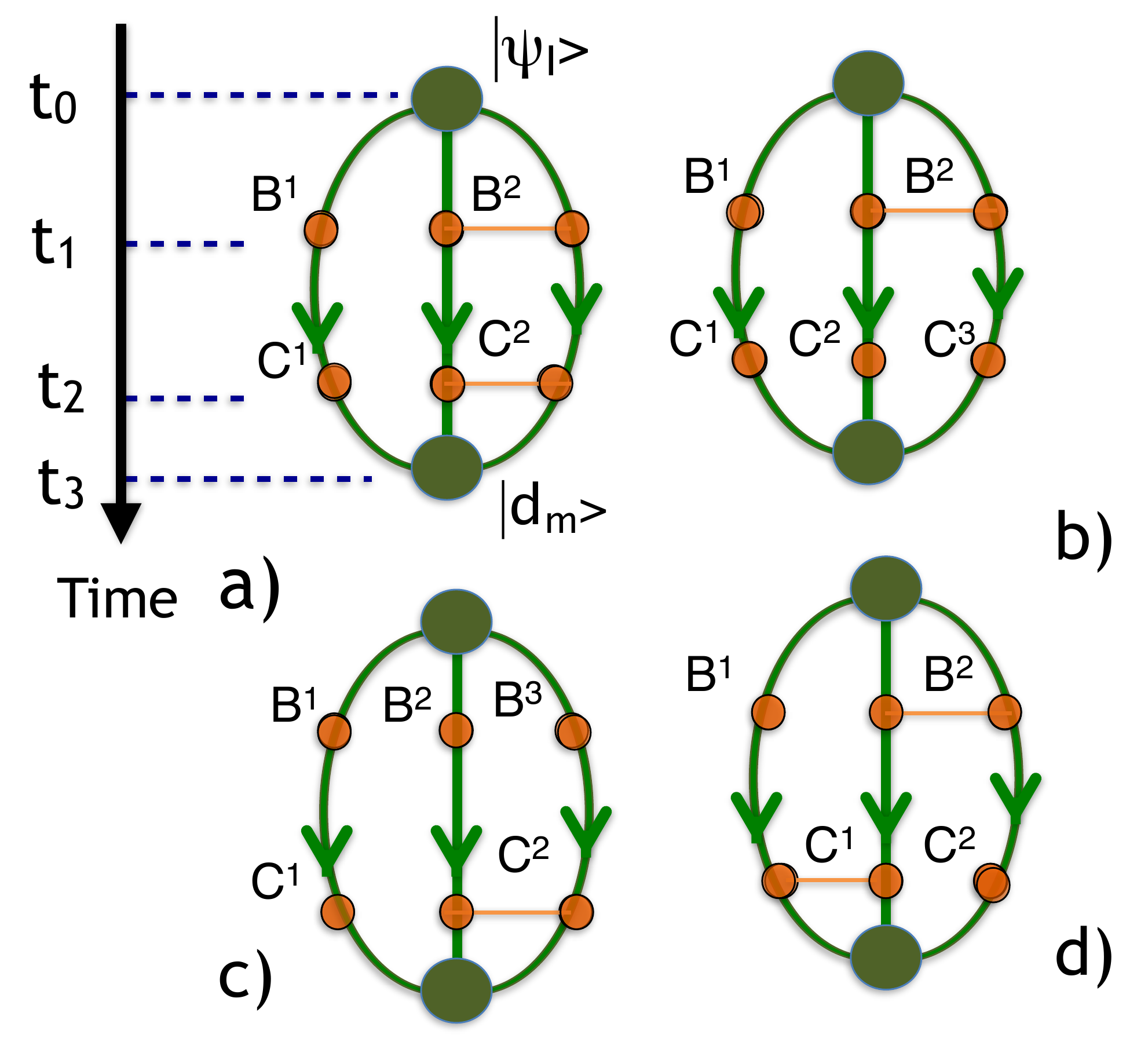}}
\caption{(Colour online) Possible choices of the eigenvalues of the operators $\hat B$ and $\hat C$, measured at $t_1$ and $t_2$. a) $B^1\ne B^2=B^3$ and $C^1\ne C^2=C^3$ allows one to distinguish only between the path $\{m,1\}$, and the union $\{m,2\cup m,3\}$.
Three other choices, b) $B^1\ne B^2=B^3$ and $C^1\ne C^2\ne C^3$; c) $B^1\ne B^2\ne B^3$ and $C^1\ne C^2= C^3$;
and d) $B^1\ne B^2= B^3$ and $C^1= C^2\ne C^3$, make all three paths, $\{m,1\}$, $\{m,2\}$, and $\{m,3\}$ "real".}
\label{fig:3}
\end{figure}
Now at $t_1<t_2<t_3$ we can measure any operator $\C$, with $C^2=C^3$, $\C=\text{diag}(C^1,C^2,C^2)$, 
without altering the ensemble, already specified by the measurements of $\B$ and $\D$ (see Fig. 3a). 
The only two possible outcomes, shown in Fig.4a will occur with the probabilities given by Eq.(\ref{b6a})
 \begin{eqnarray}\label{c4}
p(B^1,C^1,D^m)=p(B^1,D^m),\q \q m=1,2,3,\q\q\q\n
p(B^2,C^2,D^m)=p(B^2,D^m). \q\q\q\q\q\q\q\q\q\q
\end{eqnarray}
\begin{figure}
	\centering
		\includegraphics[width=8cm,height=6cm]{{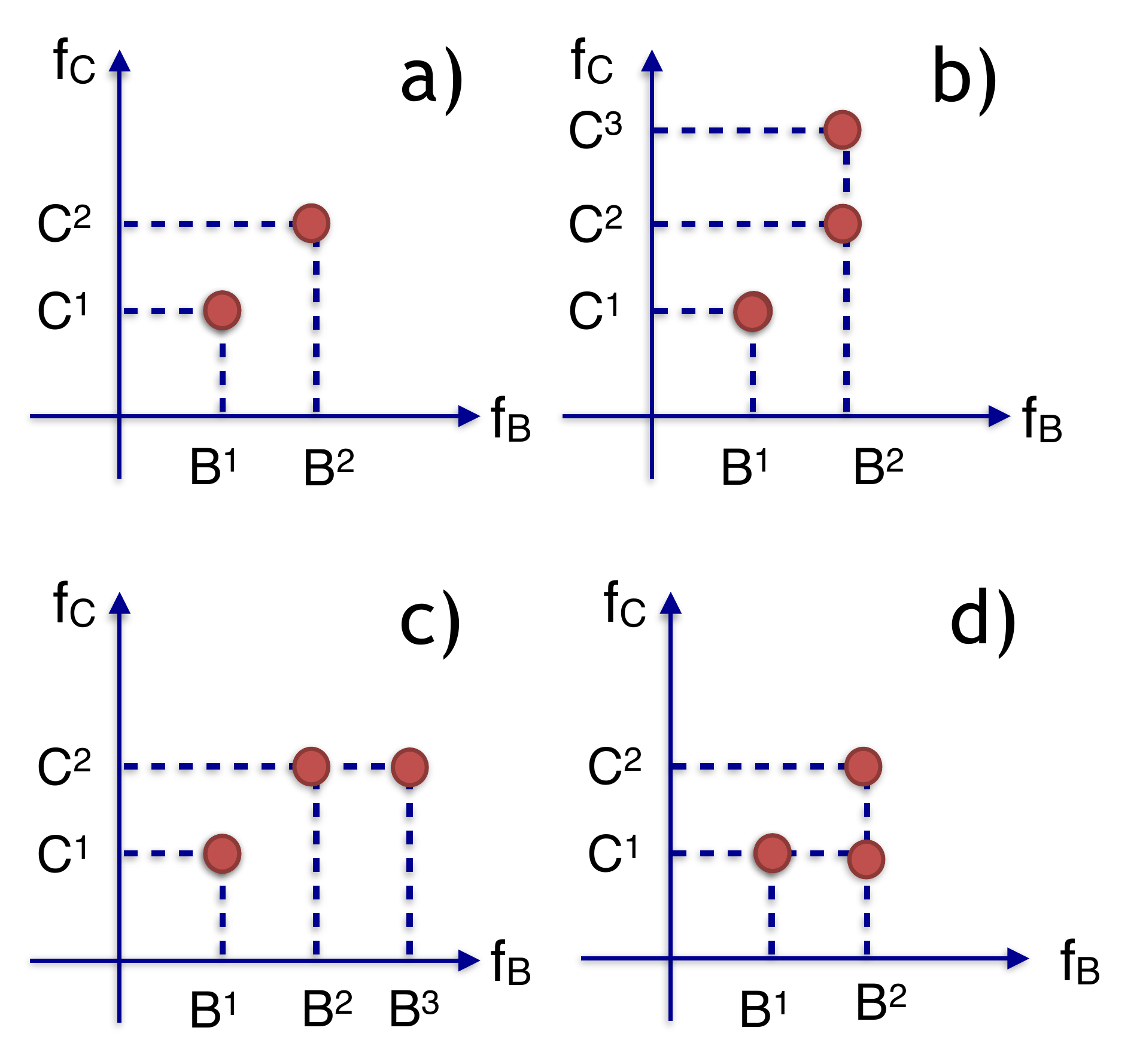}}
\caption{(Colour online) Possible readings (final pointer positions $f_B$ and $f_C$) of the meters measuring operators $\hat B$ and $\hat C$
 for the four cases shown in Fig. 3. 
}
\label{fig:4}
\end{figure}
\newline
The ensemble, however, will not remain the same if at $t_2$ we decide to measure instead a $\C$ with $C^1\ne C^2 \ne C^3$ (see Fig. 3b).
In this case, all nine virtual paths $\{m,n\}$  will be made real, and the three possible results shown in Fig.  4b will occur with the 
probabilities
 \begin{eqnarray}\label{c5}
p(B^1,C^1,D^m)=|A(m,1,|\psi_I)|^2, \q m=1,2,3,\q\q\q\n
p(B^2,C^2,D^m)=|A(m,2,|\psi_I)|^2, \q\q\q\q\q\q\q\q\q\n
p(B^2,C^3,D^m)=|A(m,3,|\psi_I)|^2. \q\q\q\q\q\q\q\q\q
\end{eqnarray}
If $\B=\text{diag}(B^1,B^2,B^3)$ is measured before $\C=\text{diag}(C^1,C^2,C^2)$ (see Fig. 3c),  the three outcomes 
shown in Fig.4c, will occur with probabilities 
 \begin{eqnarray}\label{c6}
p(B^1,C^1,D^m)=|A(m,1,|\psi_I)|^2, \q m=1,2,3,\q\q\q\n
p(B^2,C^2,D^m)=|A(m,2,|\psi_I)|^2, \q\q\q\q\q\q\q\q\q\n
p(B^3,C^2,D^m)=|A(m,3,|\psi_I)|^2. \q\q\q\q\q\q\q\q\q
\end{eqnarray}
With interference between the paths in Eq.(\ref{c1}) already destroyed by the measurement of $\B$, 
measuring of $\C$ does not modify the ensemble. 
Obtaining the net probability $p(C^2,D^m)$ for having outcomes  $C^2$ and $D^m$ reduces to ignoring the information about the value 
of $\B$, which is available in principle \cite{FeynL}. Thus, $p(C^2,D^m)=p(B^1,C^2,D^m)+p(B^2,C^2,D^m)$,
and not $|A(m,2,|\psi_I)+A(m,3,|\psi_I)|^2$, as would be the case, if $\C$ were measured on its own. 
\newline Finally, joint measurement of $\B=\text{diag}(B^1,B^2,B^2)$ at $t_1$ and $\C=\text{diag}(C^1,C^1,C^2)$,
at $t_2$ also produces nine real paths, even though, on its own, each of the two measurements is capable of producing only one pair of such paths. Indeed, as shown in Figs 3d and 4d, obtaining results $B^1$,$C^1$ and $D^m$, will confine the system to the path $\{m,1\}$, obtaining
$B^2$, $C^1$ and $D^m$ to the path $\{m,2\}$, and obtaining $B^2$, $C^2$ and $D^m$ to the path $\{m,3\}$.
\newline
A similar analysis is easily extended to systems in a Hilbert space of arbitrary number of dimensions $N$, at to intermediate measurements of more than two operators. 
\section{Non-perturbing measurements, and the reality of the "real" pathways}
Previously, we emphasised the difference between virtual paths, endowed only with probability amplitudes, and the 
real paths, to which it is also possible to ascribe probabilities.
\newline
It is usually accepted that if a measurement of an operator $\Aa$ yields the same non-degenerate eigenvalue
$A^i$ in {\it all} identical trails, than the system {\it is} (or can be said to be) in the corresponding eigenstate $|a_i\ra$.
There is a particular class of non-perturbing ("non-demolition" \cite{Real1}) measurements, which, if applied in addition to the ones being made, leave the statistical ensemble intact, 
and only confirm that the system has travelled its particular real pathway.
\newline 
As a simple example, we choose a basis $|a_k\ra$, $k=1,2,...N$,  in the Hilbert space of the system, such that it contains its initial state, 
$|\psi_I\ra=|a_1\ra$. We also choose the three measured operators to be
 \begin{eqnarray}\label{d1x}
\B=\U^{}(t_1,t_0)\hat A \U^{-1}(t_1,t_0),\q
\C=\U^{}(t_2,t_0)\hat A \U^{-1}(t_2,t_0),\q
\D=\U^{}(t_3,t_0)\hat A \U^{-1}(t_3,t_0),
\end{eqnarray}
where $\hat A$ is some operator with non-degenerate eigenvalues,
 \begin{eqnarray}\label{d2}
\hat{A}=\sum_{k=1}^N |a_k\ra A_k \la a_k|,
\end{eqnarray}
Now the results of all three measurements will aways yield $A_1$, indicating that the system 
is in the state $\U^{}(t_j,t_0)|a_1\ra$, at $t=t_j$, $j=1,2,3$.
We can add more of such measurements at times between $t_0$ and $t_3$, all confirming 
that 
the system {was} following its orbit in the Hilbert space, 
 \begin{eqnarray}\label{d3}
|\psi(t)\ra = \U(t,t_0)|\psi_I\ra.
\end{eqnarray}
Recalling the much quoted Einstein's maxim \cite{Real2}
{\it "if, without in any way disturbing a system, we can predict with certainty (i.e., with probability equal to unity) the value of a physical quantity, then there exists an element of reality corresponding to that quantity"}, 
we can also conclude that the system {\it really} was in the state $|\psi(t)\ra$ at any $t$ between $t_0$ and $t_3$.
\newline
A slightly more interesting case would involve arbitrary operators $\B$, $\C$ and $\D$ with non-degenerate eigenvalues, and a particular outcome
$(B^n,C^l, D^m)$, corresponding to a real pathway $|d_m\ra \gets |c_l\ra \gets |b_n\ra \gets |a_1\ra$. The pathway is defined only for $t=t_0, t_1,t_2$, and $t_3$, but we can fill the gaps using the procedure outlined above.
Thus, measuring
 \begin{eqnarray}\label{d4}
\Aa(t)=\sum_i |\tilde a_i(t)\ra A_i \la\tilde a_i(t)|, \q |\tilde a_i(t)\ra\equiv \U^{}(t,t_0)|a_i\ra,\n
\B(t)=\sum_i |\tilde b_i(t)\ra B_i \la\tilde b_i(t)|, \q |\tilde b_i(t)\ra\equiv \U^{}(t,t_1)|b_i\ra,\n
\C(t)=\sum_i |\tilde c_i(t)\ra C_i \la\tilde c_i(t)|, \q |\tilde c_i(t)\ra\equiv \U^{}(t,t_2)|c_i\ra,
\end{eqnarray}
for a $t_0<t<t_1$, $t_1<t<t_2$ and $t_2<t<t_3$, respectively,
will yield only the values $A^1$, $B^n$, and $C^l$, and, therefore, find the system in the states
$|\tilde a_1(t)\ra$, $|\tilde b_n(t)\ra$ and $|\tilde c_l(t)\ra$, as shown in Fig. 5.
\newline
None of these additional measurements will disturb the original statistical ensemble, and leave the probabilities $p(B^n,C^l,D^m)$ in Eq.(\ref{b6a}) unchanged. 
We will not dwell here on the problem of wave function collapse \cite{Real3}, evident from the discontinuities in the graph if Fig. 5.  It is sufficient to note that whenever the results of the measurements of $\B$, $\C$ and $\D$ are $(B^n,C^l,D^m)$
(larger dots in Fig. 5), 
we can also  ascertain (with probability equal to one, and without disturbing the system) 
that the system follows a pathway in Fig. 5, 
 \begin{eqnarray}\label{d4a}
|d_m\ra \gets |\tilde c_i(t)\ra \gets |c_l\ra \gets|\tilde b_i(t)\ra \gets |b_n\ra \gets |\tilde a_i(t)\ra \gets |a_1\ra.\q\q
\end{eqnarray}
This is the most complete description of a real path, which connects $|a_1\ra$ with $|d_m\ra$, and is produced by 
intermediate measurements of $\B$ and $\C$.
\newline
 The argument is readily extended to the case where $\B$ in Eq.(\ref{b1}) has degenerate eigenvalues. If so, obtaining $f_b=B_s$, would guarantee that for a $t_1<t<t_2$ the system is in the state
$|\psi_s(t)\ra=\U(t,t_1)\hat\pi^B_s|\U(t_1,t_0)|\psi_I\ra$, which can be verified, e.g.,  by measuring the projector
$|\psi_s(t)\ra\la \psi_s(t)|$. 
\begin{figure}
	\centering
		\includegraphics[width=8cm,height=4cm]{{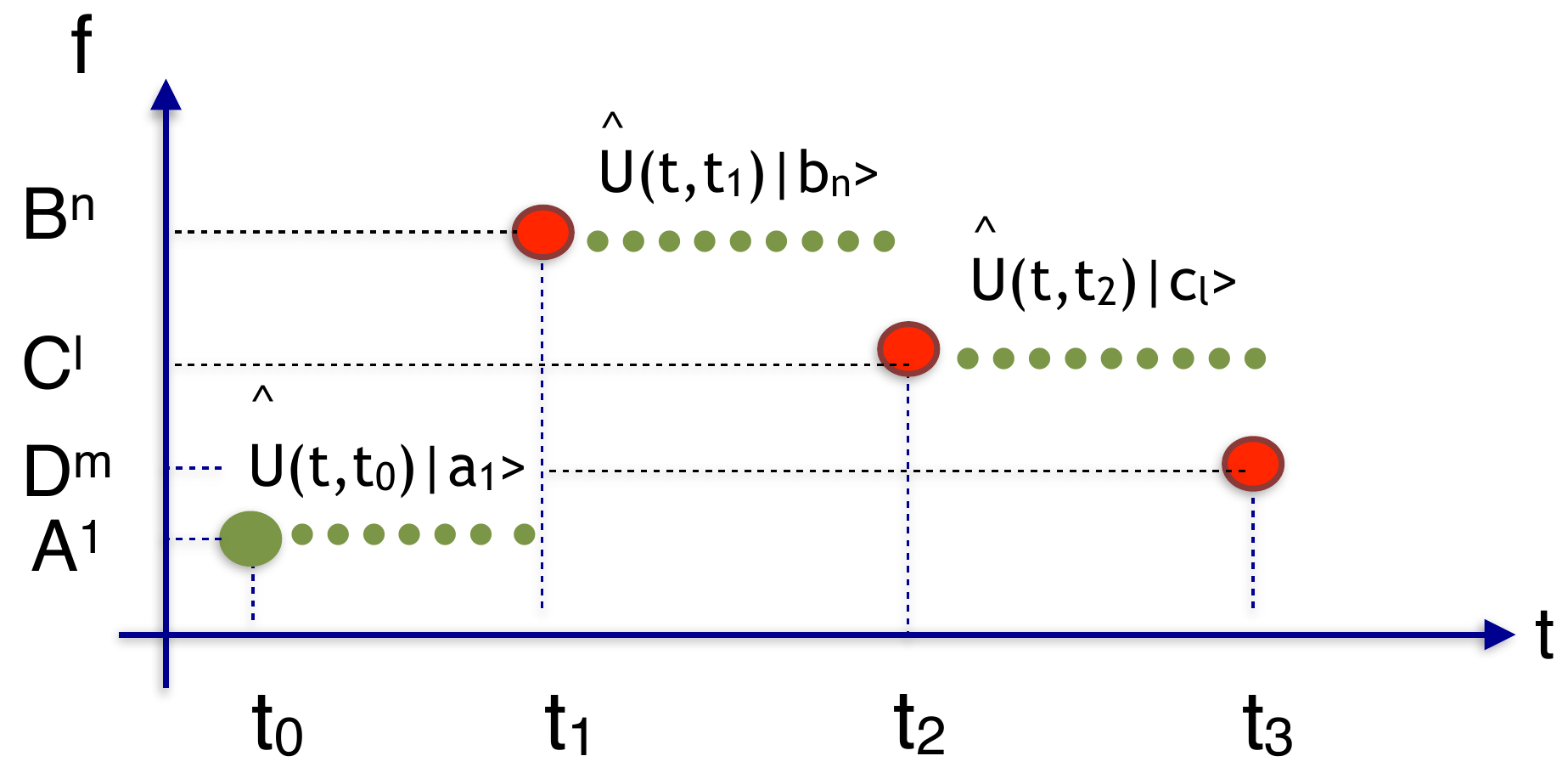}}
\caption{(Colour online) The system is prepared in a state $|\Psi_I\ra=|a_1\ra$, and three non-commuting 
operators $\B$, $\C$ and $\D$ are measured, with the outcomes $B^n$, $C^l$ and $D^m$, respectively (red dots).
Adding intermediate measurements of the operators in Eq.(\ref{d4}) leaves the probabilities $p(B^n,C^l,D^m)$ 
unchanged. 
Without disturbing the system,
one is able to verify operationally that
between two ensemble-forming measurements, 
 the system continues in a state, produced by the most recent collapse of its wave function.}
\label{fig:5}
\end{figure}
\newline
In summary, additional measurements, which can be performed on an existing ensemble fall into two categories . 
Perturbing, or {\it ensemble forming}, measurements produce new real pathways,
 and {fabricate} a different ensemble, in general completely distinct from the one we had before.
Such measurements do not occur in a classical theory, where it is always assumed that observations should have no effect on the monitored system.
{\it Non-perturbing} measurements, on the other hand, allow one to confirm additional properties of the system \cite{Real1}, without modifying the existing ensemble. 
With respect to such measurements, the situation is similar to the one in classical physics, where the system's evolution, interrupted by random events, 
can be monitored without altering its course.
Next we take a closer look at some of the "quantum paradoxes", recently discussed in the literature.

\section{The "three-box paradox"}
The simplest example of a quantum mechanical  "paradox" is the three-box case \cite{3Ba} - \cite{3Bc}. 
A $3$-state system, similar to the one discussed in Sect.VI, can reach a final state, $|d_1\ra$ from some $|\psi_I\ra$, 
by passing through one of the orthogonal states,  $|b_n\ra$, $n=1,2,3$.
There are three virtual paths,
 \begin{eqnarray}\label{d0}
\{n\}  \q \text{via} \q |d_1\ra \gets |b_n\ra \gets|\psi_I\ra, \q n=1,2,3,
\end{eqnarray}
and three probability amplitudes (for simplicity we put $\HH\equiv 0$) $A(n)\equiv A(1,n|\psi_I)=\la d_1|n\ra \la n|\psi_i\ra$.
Allegedly, a paradox arises if the initial, $|\psi_I\ra$,  final, $|d_1\ra$, and intermediate, $|b_n\ra$,  states
are chosen in such a way that 
 \begin{eqnarray}\label{d1}
A(1)=-A(2)=
 A(3)\equiv A \ne 0.
\end{eqnarray}
 \begin{figure}
	\centering
		\includegraphics[width=9cm,height=5cm]{{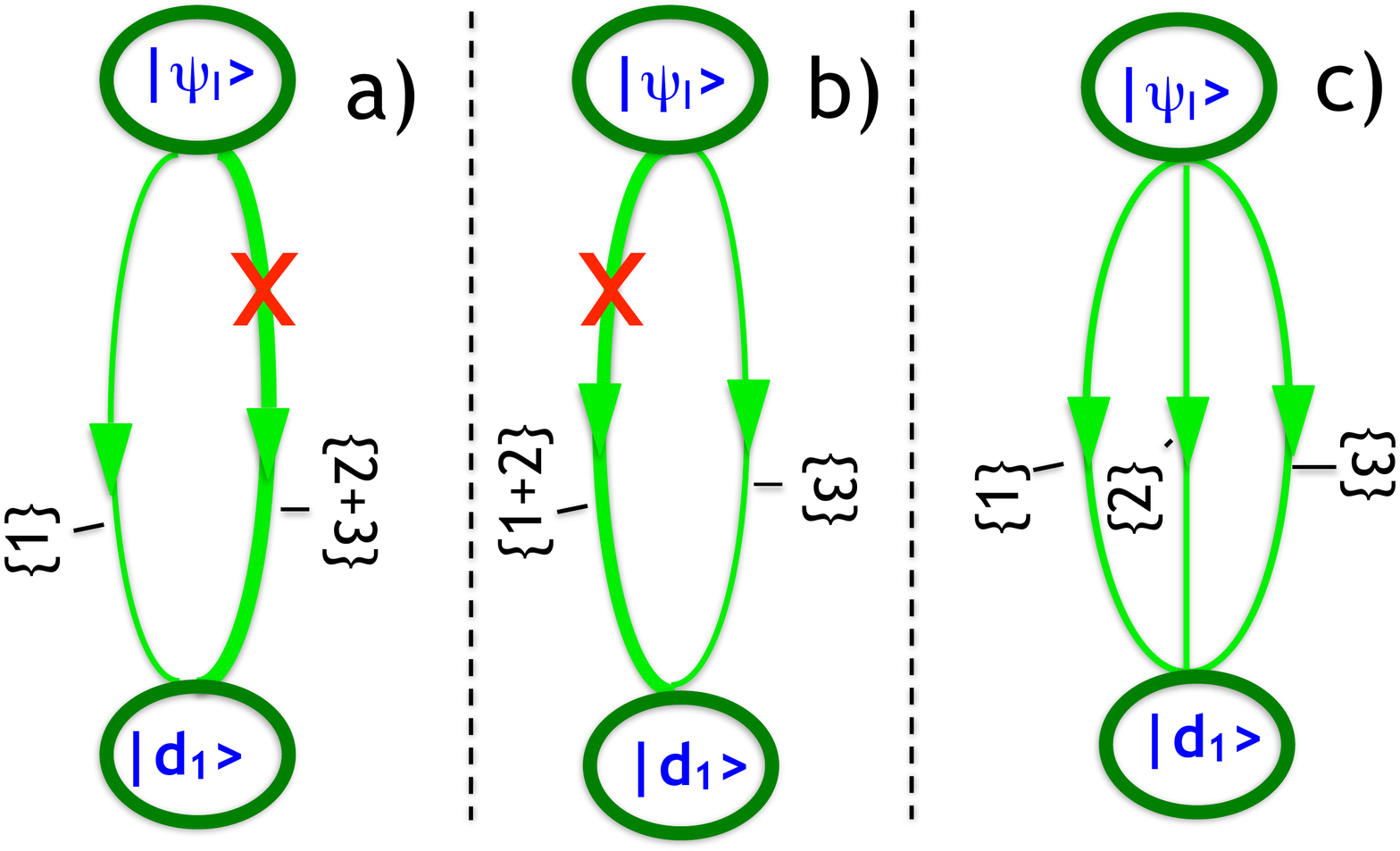}}
\caption{(Colour online) Real pathways produced from the virtual paths in Eq.(\ref{d0}) by measurement
of an operator  a) $\B=\text{diag}(B^1,B^2,B^2)$; b) $\C=\text{diag}(C^1,C^1,C^3)$, and c)
$\hat F=(F^1,F^2,F^3)$. Pathways, for which the probability happens to be zero, are marked by a cross.}
\label{fig:6}
\end{figure}
\noindent
With no measurements made, the system arrives at $|d_1\ra$ with a probability $P(D^1)=|A|^2$.
A measurement of an operator $\B=\text{diag}(B^1,B^2,B^2)$, leaves the probability $P(D^1)$ unchanged, 
and {\it always} yields the value $B^1$. According to the "eigenvalue-eigenstate correspondence"
described at the beginning of Sect. VII, this means that the system always takes the path $\{1\}$. 
By the same token, measurement of an operator $\C=\text{diag}(C^1,C^1,C^3)$ will reveal that 
the system always takes the path $\{3\}$. 
\newline
The authors of \cite{3Bb} suggest that {\it "a paradox . . . is that at a particular time . . . a particle is in some sense both with certainty in one box, and with certainty in another box"}. The claim is, however, unwarranted. 
Measurements of $\B$ and $\C$, both create ensembles with two real paths (see Figs. 6a and 6b), but the ensembles are incompatible. As in Sect III., there are no probabilities $p_1$ and $p_2$, which would yield correct average values for both 
 $\B$ and $\C$, since equations 
 \begin{eqnarray}\label{d1a}
B^1p_1+B^2 p_2=B_1, \q C^1 p_1+C^3p_2=C^3,\q p_1+p_2=1 \q\q
\end{eqnarray}    
have no solutions. The system does take paths  $\{1\}$ and  $\{3\}$   with certainty, but this happens in two different circumstances, 
created by different measurements.
\newline
We can illustrate this by employing non-perturbing measurements of Sect. VII. In the first case, non-perturbing would be the measurements of 
projectors $|\psi_I\ra\la \psi_I|$, and $\hat{\pi}(2,3)=|b_2\ra\la b_2|+|b_3\ra\la b_3|$, before and after the measurement of $\B$, 
respectively. They would confirm that at all times the system follows a path $|d_1\ra \gets |b_1\ra \gets |\psi_I\ra$.
Similar measurements of $|\psi_I\ra\la \psi_I|$, and $\hat{\pi}(1,2)=|b_1\ra\la b_1|+|b_2\ra\la b_2|$, 
would confirm that the system always follows different path, $|d_1\ra \gets |b_3\ra \gets |\psi_I\ra$, if $\C$ is measured. 
In both cases, an attempt to probe the structure of the real paths further, e.g., by measuring also an operator
$\hat F=(F^1,F^2,F^3)$ with all different eigenvalues, would destroy the existing ensemble, create three real paths, 
shown in Fig. 6c, and increase the probability to arrive at $|d_1\ra$  from $|A|^2$ to $3|A|^2$.
\newline
In summary, the authors of \cite{3Ba} - \cite{3Bc} compare the results corresponding to {\it different}
physical conditions, and have no reason to claim a paradox. A person can be certainly at work on a Monday, or at home if this Monday turns out to be a holiday. Since each Monday is ether a workday, or a holiday, there is no need 
to be in two places at the same time.
\section{Photon's past in the "nested interferometer"}
Our next example is a network of optical fibres and beam splitters ($BS$) shown in Fig. 7a.
At the point $In$, a photon in injected in a wave packet state $|\psi_I\ra$ and ends up being split into three final states, $|d_m\ra$, $m=1,2,3$, leaving system in three different ways, as shown in Fig. 7. 
The system is tuned in such a way that no part of the wave function travels through the fibre $F$ (dashed), and we are interested in detecting the photon 
by means of a detector $D$. The subject of the ongoing discussion, often touching on the topic  of "weak measurements" (WM), is the past of the detected photon  \cite{Nest1}-\cite{Nest6}. We will briefly return to the WM in Sect. XIII, and study first  what would happen 
if its presence in one of the branches of the interferometer were ascertained by means of accurate "strong" intermediate measurements. 
\newline
In the absence of such measurements there are three virtual paths leading to the desired outcome,
  \begin{eqnarray}\label{e1}
\{1\} \q \text{via}\q D\gets F \gets A \gets E \gets In, \n
\{2\} \q \text{via}\q D\gets F \gets B \gets E \gets In, \n
\{3\} \q \text{via}\q D\gets C \gets In, \q\q\q\q
\end{eqnarray}
\begin{figure}
	\centering
		\includegraphics[width=9cm,height=5cm]{{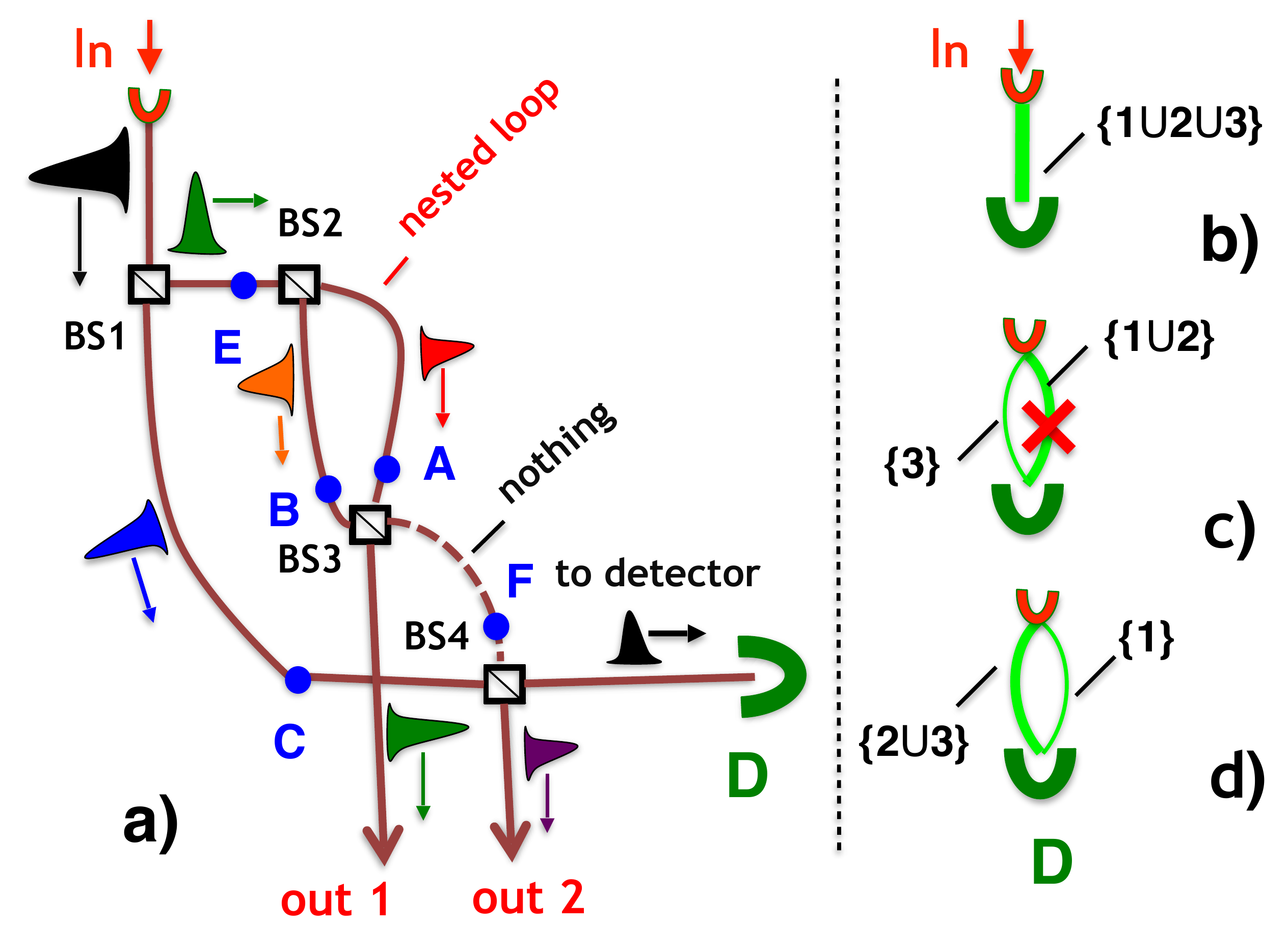}}
\caption{(Colour online) a) Schematic of an interferometer, containing an additional
loop, nested between the beamsplitters BS2 and BS3. 
Having entered the system, the photon can leave it in three different ways: $\{out1,\q out2, \q to\q detector\}$.
Also shown the real pathways, leading to the detector: b) with no measurements made; c) with $\B$ in Eq.(\ref{e2a}) measured at $E$;
and  d) with $\C$ in Eq.(\ref{e2b}) measured and $A$}
\label{fig:7}
\end{figure}
\newline
with the probability amplitudes $A(i) \equiv A(D,i|In)$, $i=1,2,3$,  and
we can proceed as in the previous Section.
Paths $\{1\}$ and $\{2\}$ are merged together before the  $BS2$ and after  the $BS3$, and we 
can describe detection of the photon at $E$ as a measurement of an operator 
\begin{eqnarray}\label{e2a}
\B=\text{diag}(1,1,0),
\end{eqnarray}
which does not distinguish between these paths. Detection of a photon at the point $A$ 
may then 
correspond to measuring an operator
\begin{eqnarray}\label{e2b}
\C=\text{diag}(1,0,0). 
\end{eqnarray}
To ensure that no part of the wave function passes through $F$, we must have
\begin{eqnarray}\label{e2}
A(1)=-A(2) \ne 0,
\end{eqnarray}
together with  $A(3) \ne 0$, which would allow  $D$ to click occasionally.
Then, an attempt to determine whether the photon passes through $E$ would create an ensemble with two real pathways,
the path $\{3\}$, travelled with a probability $P(via \q C) \equiv P(D,C)=|A(3)|^2$, and
the union (superposition) $\{1\cup 2\}$,  never travelled, since $P( not\q C)\equiv P(D,E)=|A(1)+A(2)|^2=0$.
Note that we can add another such meter at $F$, which would not perturb the system, and merely confirm what has just been said.
A similar situation is shown if Fig. 3a.
\newline
Measuring at $A$ the operator $\C$ in Eq.(\ref{e2b})
 would create real pathways $\{1\}$, and $\{2\cup3\}$ travelled with the probabilities
$P(via \q A)=|A(1)|^2\ne 0$ and $P( not\q A)=|A(2)+A(3)|^2\ne 0$.
 The same will be true if we try to detect the photon at $B$, by measuring a $\C'=\text{diag}(0,1,0)$.
 The photon 
will be found there with a probability $P(via \q B)=|A(2)|^2=|A(1)|^2\ne 0$, and detected at $D$ with a probability
$|A(2)|^2+|A(1)+A(3)|^2\ne 0$. 
\newline
Finally, installing meters both at $E$ and $A$,  will turn all three paths (\ref{e1}) into real pathways, 
labelled by three possible outcomes (cf. Figs. 3d and 4d) 
as $(via\q E, via\q A)$, $(via\q E, not\q A)$ and 
$(not\q E, not\q A)$, and travelled with the probabilities $|A(1)|^2$, $|A(2)|^2$ and $|A(3)|^2$, respectively.
The same effect can also be achieved in a different way, for example, by placing meters at both $E$ and $B$, 
or at both $A$ and $B$.
\newline
One may (although by no means needs to) find it strange that the meters at $E$ and $F$ see no photons, 
whereas a meter at $A$ or $B$ can find one there. This could be elevated to a "paradox" of finding that the photons
{\it "...have been inside the nested interferometer [loop]..., but they never entered and never left the nested interferometer..."} \cite{Nest2}, but only if this happened within the same statistical ensemble,
i.e., if it were possible o assign non-negative probabilities to the three paths in Eqs.(\ref{e1}). 
However, the compatibility test fails, just as in the previous Section, and indicates that the conflicting results correspond to two different ensembles, produced by different sets of measurements. There is no contradiction between having a room full of people, and a room no one can possibly enter, as long as these are two different rooms.
\section {Interaction-free measurements and the double-slit experiment}
So far, we have looked at the examples where the added intermediate measurements destroyed interference between virtual paths, 
or recombined them, to produce different real pathways. Another way to create a different ensemble would be to provide 
new final states for the system, and redirect to them some of the existing paths. 
Consider, for example, the set up in Fig. 7a consisting of two fibres, two beamsplitters, $ BS1$ and $BS2$, and three detectors, $D1$, $D2$ and $D3$. There is a switch ($SW$) which, if set to position $b$, would redirect towards detector $D3$ a photon, otherwise destined for $BS2$. The  system is tuned is such a way that with  $SW$ set to $a$, the photon always arrives in $D1$, and never in $D2$. 
With the switch in this position, there are two virtual paths, 
\begin{eqnarray}\label{f1ax}
\{ 1\} \q \text{via}\q D1\gets A \gets In,\n 
\{ 2\}\q \text{via}\q  D1\gets B \gets In, 
\end{eqnarray}
leading to $D1$, and another two, 
\begin{eqnarray}\label{f1b}
\{ 3\}\q \text{via}\q  D2\gets A \gets In,\n 
\{ 4\}\q \text{via}\q  D2\gets B \gets In, 
\end{eqnarray}
leading to $D2$.
The corresponding amplitudes are $A(1) \equiv A(D1,A|In)$, $A(2)\equiv A(D1,B|In)$, $A(3)\equiv A(D2,A|In)$, and $A(4)\equiv A(D2,B|In)$. 
Two real pathways, $\{ 1\cup 2\}$ and $\{ 3\cup 4\}$, travelled with the probabilities $|A(1)+A(2)|^2$ and 
$|A(3)+A(4)|^2$, lead to two alternative outcomes (see Fig. 8b).  To ensure that $D2$ never clicks, we may choose
\begin{eqnarray}\label{f1}
A(3)=-A(4) \ne 0,
\end{eqnarray}
one particular choice being  $A(1)=i/2$, $A(2)=i/2$, $A(3)=1/2$ and $A(4)=-1/2$ \cite{IFMsplit}.
\newline
With the switch in position $b$, there are three possible destinations, $D1$, $D2$ and $D3$, reached via real pathways, 
\begin{eqnarray}\label{f1a}
\{ 1\}\q \text{via}\q  D1\gets A \gets In,\n
 \{ 2\}\q \text{via}\q D2\gets A \gets In, \n 
\{ 3\}\q \text{via}\q  D3 \gets B \gets In. 
\end{eqnarray}
with amplitudes $A(1)=i/2$, $A(2)=i/2$, and $A(3)=i/\sqrt{2}$,
travelled with probabilities 
$|A(1)|^2=1/4$, $|A(2)|^2=1/4$ and $|A(3)|^2=1/2$, respectively (see Fig. 8b).
\begin{figure}
	\centering
		\includegraphics[width=9.5cm,height=5cm]{{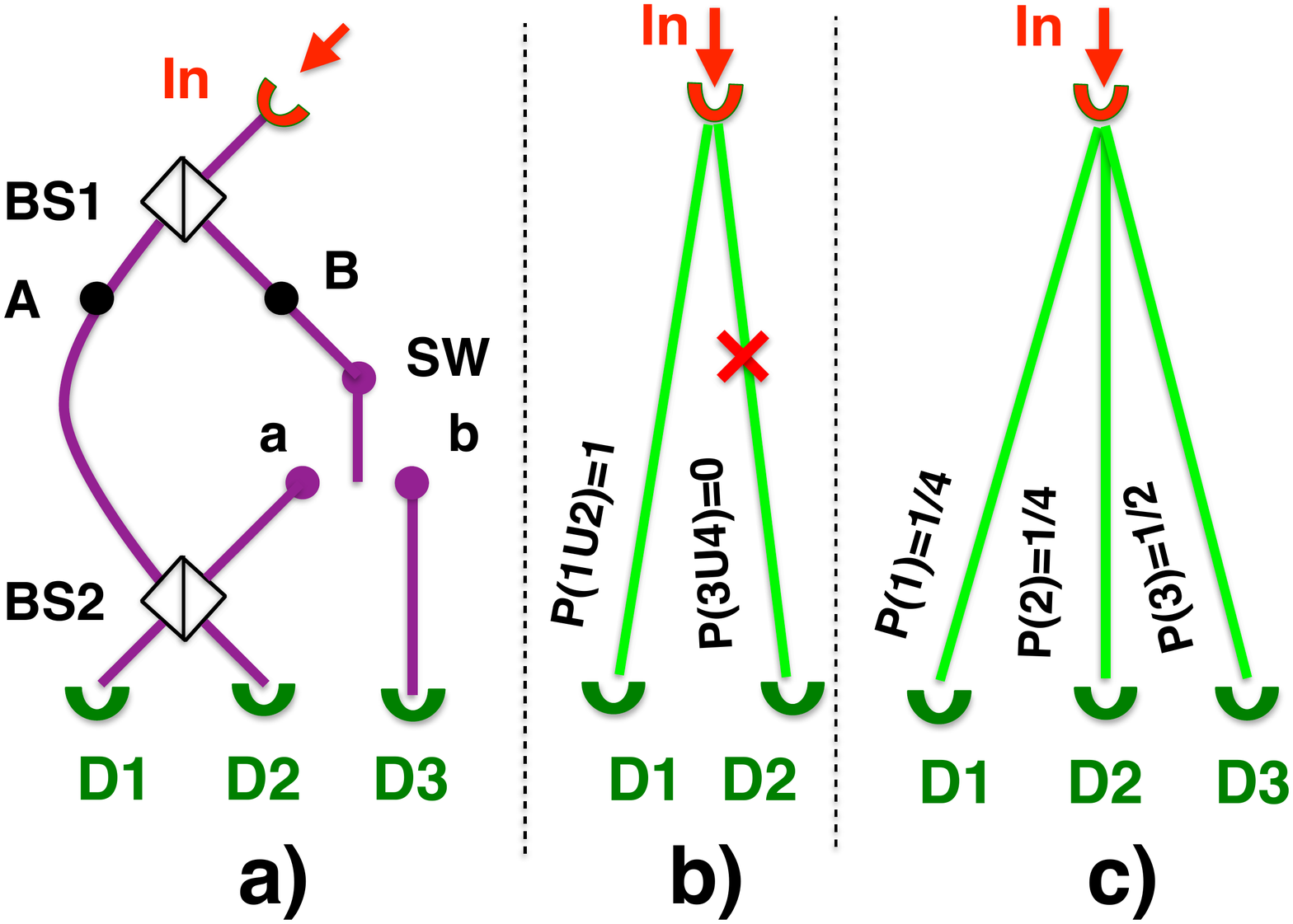}}
\caption{(Colour online) a) A system of fibres, beam splitters (BS) and detectors (D).
A wave packet, introduced at $In$, is split in two parts at $BS1$.  One part always goes to $BS2$, while the other  
can be directed to $BS2$, or to the detector $D3$, depending on the position of the switch $SW$.
b) The real paths of the photon with $SW$ set to "a". c) The real paths of the photon with $SW$ set to "b".
}
\label{fig:9}
\end{figure}
\newline
An effect, similar to the one caused by setting the switch to $b$, can be achieved by replacing the switch with an additional system 
(a "bomb" in the more colourful language of Eliezur and Vaidman \cite{IFM1}) whose internal state would certainly 
change (the bomb would explode) if the photon passes through $B$ in Fig. 8a.
This would also affect the destructive interference which prevents the photons from reaching $D2$, 
and allow the detector $D2$ to click occasionally. 
What could be considered a "paradoxical" aspect  of these experiments, is that the photon can now arrive at $D2$ only through the left arm of the setup in Fig. 8a,
and is, therefore, capable of detecting the "bomb" in the other arm without interacting with it. Hence, 
the term "interaction free measurement" often used in the literature, 
{for example, in \cite{IFM2}), where a clever application of the Zeno effect allowed the authors to make the maesurement almost 
$100\%$ efficient.}
\newline
The effect itself is not new. One might as well consider a detector placed at the dark (no electrons) fringe of a
Young's double slit experiment, which starts detecting electrons if one of the slits is blocked. As early as in 1935
Bohr used this example to argue that, in the presence of interference, the idea of a particle passing through 
one of the holes must be wrong. In \cite{IFMBohr} one reads
{\it "If we only imagine the possibility that without disturbing the phenomena we determine through which hole the electron passes, we would truly find ourselves in irrational territory, for this would put us in a situation in which an electron, which might be said to pass through this hole, would be affected by the circumstance of whether this [other] hole was open or closed..."}
Similarly one may wonder
 how can the photon, arriving at $D2$, know that $SW$ is set to $b$, if it has never paid a visit there?
\newline
Here we want to look at the problem from a slightly different angle, namely, by considering a classical setup
performing the same function as the quantum one, and try to pinpoint the principle differences between the two.
A classical model shown in Fig. 9 consist of conduits, directing a classical particle (a ball) inserted from the top, towards one of the three receptacles $D1$, $D2$ and $D3$. 
The triangles in Fig. 9 represent branching connectors, $(Br)$ which, if set in a state 
$(p_L,p_R)$, randomly direct a ball, introduced from above, to the left and right conduits,  with adjustable probabilities $p_L$ and $p_R=1-p_L$, respectively. The top connector is always set to $(1/2,1/2)$, and there is also a switch $(SW)$, whose function is similar to that of the switch in Fig. 8a.
The statistics will be collected by counting the balls ending up in different receptacles $D1$, $D2$, and $D3$. 
\newline
With the switch in Fig. 9 set to $a$, and $Br1$ and $Br2$ both set to $(1,0)$ the classical statistics will be the same 
as in the quantum case in Fig. 8a with the $SW$ set to $a$.  With the switch in Fig. 8 changed to $b$, {\it and} $Br1$  set to $(1/2,1/2)$, we reproduce the quantum case with $SW$ set to $b$. 
\newline
The similarities are obvious. In both cases we much change configuration of the setup. There is no problem with a classical ball rolling into $D2$ without consulting first 
the position of the switch. Arguably, there should not be one with the electron either.
\newline
The principle difference is in {\it how} the changes can be made. While in the quantum case
single operation on the switch is sufficient, in the classical case we much effect changes in two different, possibly distant, parts of the setup, $SW$ and $Br1$. Thus, quantum mechanics allows one to achieve with a single local operation, what classically requires at least two operations in different places. This is a consequence of quantum non-locality, which is well known to forbid considering individual  states of an entangled pair \cite{IFMent}. In the same sense, it forbids separate consideration of the branches of the interferometer in Fig. 8a, through which individual parts of the photon's wave packet 
must propagate before the photon reaches its destination.
\begin{figure} 
	\centering
		\includegraphics[width=8cm,height=6cm]{{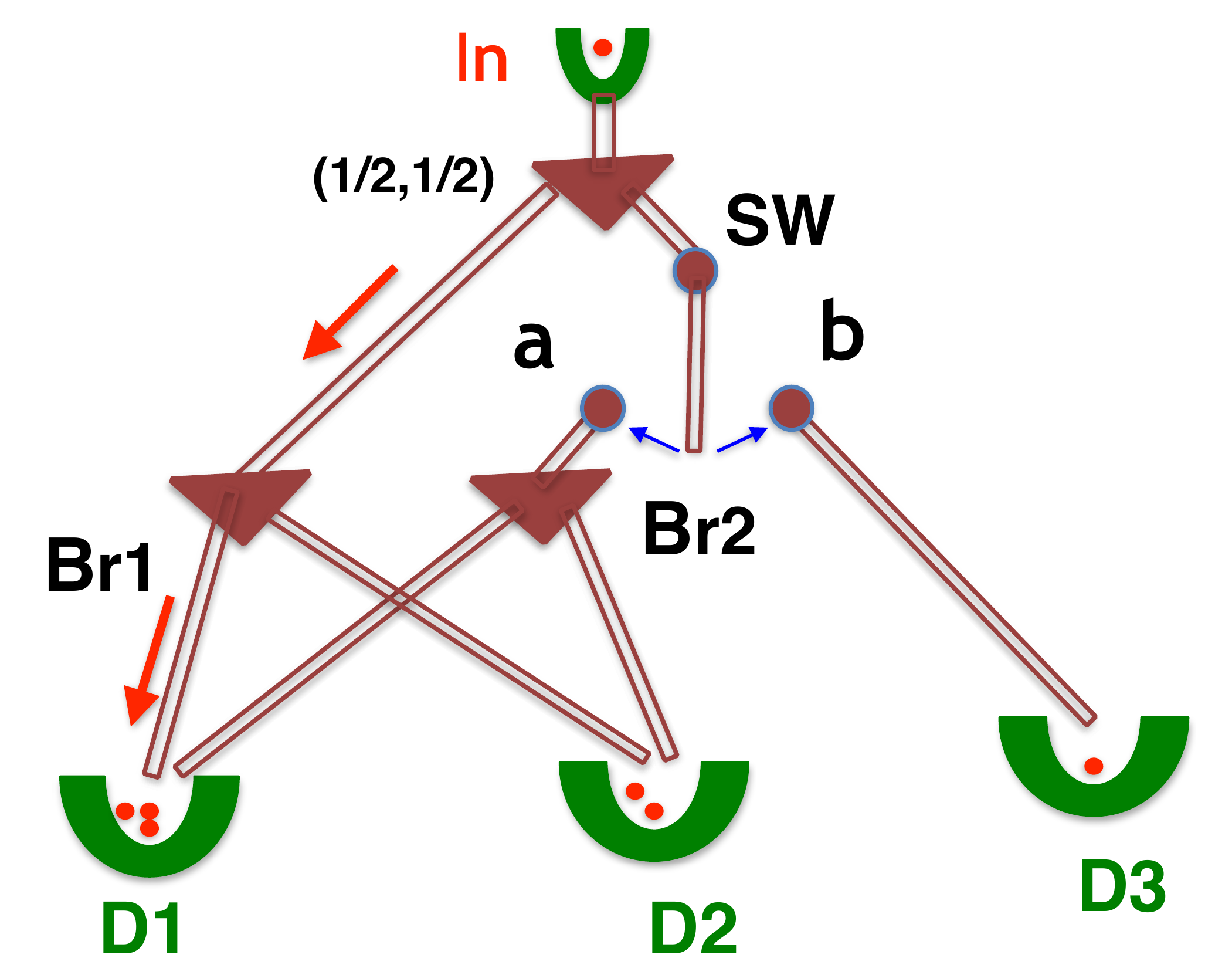}}
\caption{(Colour online) A classical analogue of the quantum setup shown in Fig. 8a,
consisting of conduits (double lines)  and branching connectors ($Br$, triangles).
After entering a connector, classical particles can exit to the left or to the right, with known probabilities $p_L$ and $p_R=1-p_L$, respectively. 
A switch ($SW$) in position "a" directs the particle to $Br2$ and, in position "b",
towards the receptacle $D3$. The probability of a particle, introduced  from the top ($In$), 
ending up in one of the receptacles is the same, as that of a quantum particle in Fig. 8a being registered by the corresponding detector.}
\label{fig:10}
\end{figure}
\section {Composite systems. The Hardy's "paradox"}
Analysis of Sect. V remains valid when we deal with a composite system $S$, consisting
of two sub-systems, $S'$ and $S''$, which exist  in $N'$ and $N''$-dimensional Hilbert spaces, respectively.
There are $N=N'\times N''$ basis states, e.g., direct products each sub-system's basis functions, and 
a typical operator $\B=\sum_{i=1}^N |b_i\ra Bî\la b_i|$ is represented by an $N\times N$ matrix.
\newline 
The setup which could be used for observing the so-called "Hardy's paradox" \cite{HARDY1, HARDY2, HARDY3, HARDY4} is shown in Fig.  10a.
It consists of two systems, each containing two wave guides, two $50:50$ beam splitters, and two detectors.
The inner waveguides approach each other at $F$ in such a way that, although unable to jump between the waveguides,
the particles may interact. 
Thus, the presence of one particle near $F$ in one of the wave guides, may block the passage of the remaining particle through the other wave guide. 
For the sake of the argument, one can assume that if a particle and its anti-particle are injected into the left and right subsystem, their encounter 
in the vicinity of  $F$ would result in mutual annihilation, in which case none of the detectors will click, 
and a pair of $\gamma$ quanta will be observed instead. 
The system is tuned in such a way that injection of two non-interacting particles always results in the detections 
by detectors in $D1$ and $D3$, while $D2$ and $D4$ never click. 
The property of interest is that, with the annihilation possible, there is finite probability for $D2$ and $D4$ to click simultaneously. This cannot be a result of annihilation, because there would be no particles to arrive 
at $D2$ and $D4$. Yet, without the possibility of annihilation, the two detectors would not be able to click.
\newline
Our detailed analysis of possible measurements in the Hardy's
setup can be found in \cite{HARDY4}, and here we limit ourselves to a much shorter discussion of its essential features.
Without annihilation, 
there are, in principle,  four possible outcomes which we denote 
$(D1,D3)$, $(D1,D4)$, $(D2,D3)$ and $(D2,D4)$, each of which can be reached from the initial state 
of two non-interacting particles, $|\Psi_I\ra=|\psi'_I\ra|\psi''_I\ra$, via four virtual paths. For example,  four such paths lead to the outcome of interest, $(D2,D4)$  [$(A,C)$ indicates that the first and the second particles pass through
$A$ and $C$, etc.], 
\begin{eqnarray}\label{h1}
\{ 1\}\q \text{via}\q  (D2,D4) \gets (A,C) \gets I,\n
\{ 2\}\q \text{via}\q  (D2,D4) \gets (A,D) \gets I,\n
\{ 3\}\q \text{via}\q  (D2,D4) \gets (B,D) \gets I,\n
\{ 4\}\q \text{via}\q  (D2,D4) \gets (B,C) \gets I.\n
\end{eqnarray}
Together they form a single real pathway, travelled with a probability
\begin{eqnarray}\label{h4x}
P(1\cup 2 \cup 3\cup 4) =|A(1)+A(2)+A(3)+A(4)|^2 =0.\q
\end{eqnarray}
Given the symmetry of the setup in Fig. 10a, and the fact that $D2$ and $D4$ never click together, for the amplitudes $A(i)$ 
we should have
\begin{eqnarray}\label{h2}
A(1)=-A(2)=A(3)=-A(4)=A \ne 0, 
\end{eqnarray}
where $A$ is some complex number. 
\begin{figure} 
	\centering
		\includegraphics[width=8cm,height=6cm]{{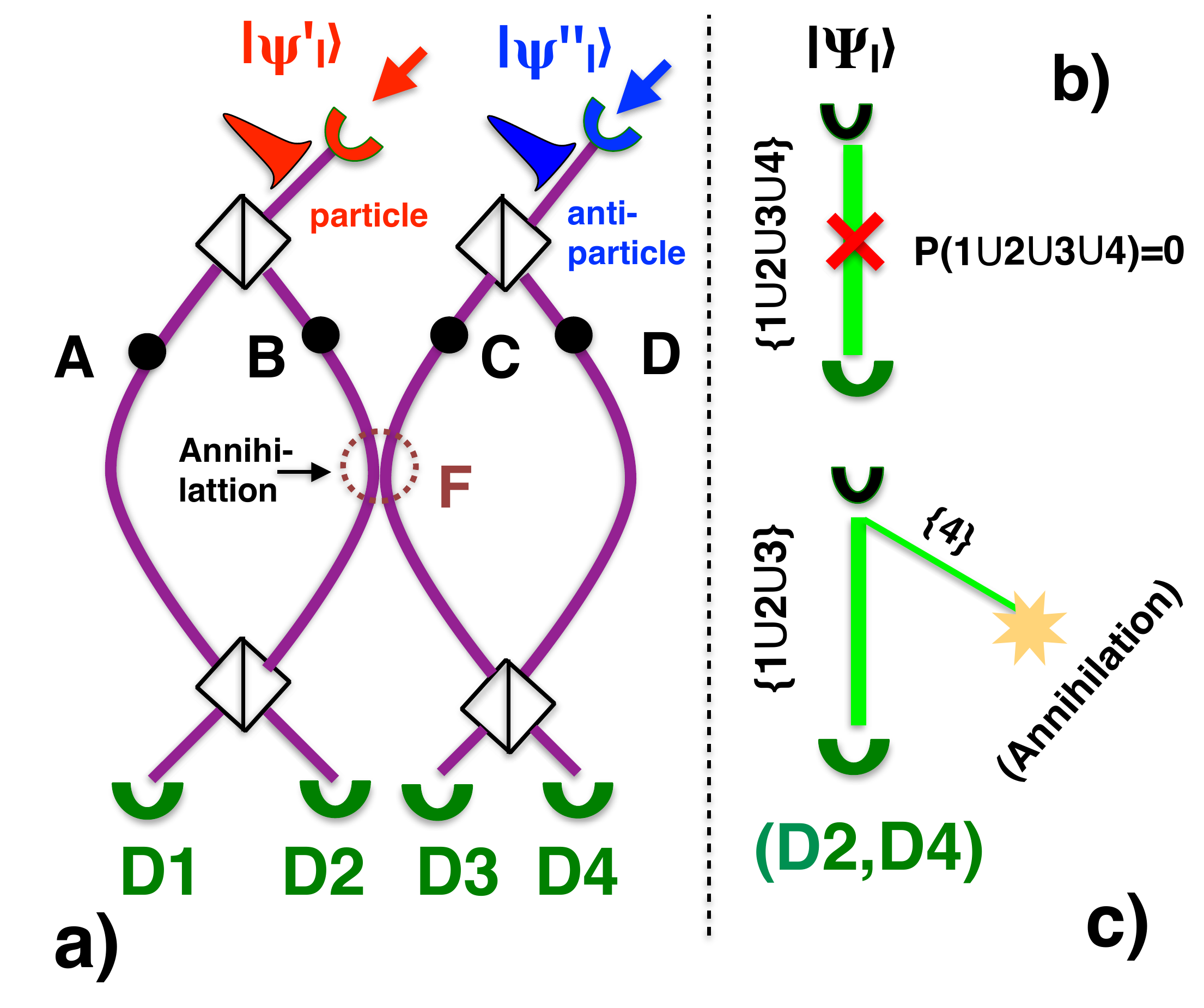}}
\caption{(Colour online) a) A system of wave guides and beam splitters, in which two particles can reach detectors $D1$, $D2$, and $D3$, and $D4$, respectively. The system is tuned in such a way that,  if the particles 
do not interact, only detectors $D1$ and $D3$ can click. Allowing the particles to annihilate, if they meet 
at $F$, results in all pairs of detectors, $(D1,D3)$, $(D1,D4)$, $(D2,D3)$, and $(D2,D4)$, clicking with non-zero probabilities.
b) Single real pathway leading to the outcome $(D2,D4)$ for a pair of independent particles.
c) The real pathways leading to $(D2,D4)$, and to annihilation, for a particle-antiparticle pair.}
\label{fig:11}
\end{figure}
\newline
With the annihilation possible, there are only three virtual paths leading to the outcome $(D2,D4)$, 
\begin{eqnarray}\label{h3}
\{ 1\}\q \text{via}\q  (D2,D4) \gets (A,C) \gets I,\n
\{ 2\}\q \text{via}\q  (D2,D4) \gets (B,D) \gets I,\n
\{ 3\}\q \text{via}\q  (D2,D4) \gets (A,D) \gets I,
\end{eqnarray}
which together form a single real pathways travelled with a probability
\begin{eqnarray}\label{h4}
P(D2,D4)=P(1\cup 2 \cup 3) =\q\q\q\n
|A(1)+A(2)+A(3)|^2 =|(A(1)|^2 =|A|^2.\q
\end{eqnarray} 
The fourth path now leads to a different observable outcome, 
\begin{eqnarray}\label{h5}
\{ 4\}\q \text{via}\q  (Annihilation) \gets (B,C) \gets I.\n
\end{eqnarray}
and is a real pathways in its own right.
\newline
Note the similarity with the case of interaction-free measurement, discussed in the previous Section.
As in Sect. X, one of the paths of the composite system (particle-antiparticle pair) is diverted to a different 
destination (annihilation), and becomes a new real pathway.
Three remaining virtual paths continue to form a real pathway, but their amplitudes no longer add up to zero, 
so that $D2$ and $D4$ are able to click in unison.
\newline 
It is easy to evaluate the probability of joint detection vy $D2$ and $D4$ from general considerations.
Let $Y$ be the amplitude for the first particle to reach $D1$ via 
$D1\gets A \gets |\psi'_I\ra$, and 
 by symmetry, also the amplitude to reach $D1$ via $B$. 
 Since $D1$ always clicks, if only the first particle is introduced, 
 we have $2Y=1$, or $Y=1/2$.
 By symmetry, $Y$ is also the amplitude 
for the second particle to reach $D3$ via $C$. 
The amplitudes
for independent events multiply \cite{FeynL}, so that $A(1)=Y^2=1/4$. 
Thus, the possibility annihilation allows $D2$ and $D4$ to click simultaneously.
From Eq. (\ref{h4}), the probability of this to happen is $P(D2,D4)=1/16$.
\newline
It is a simple matter to evaluate also the probabilities for the remaining four outcomes.
With the diversion of the fourth path to annihilation, the amplitude for $D1$ and $D3$ to click 
is reduced by one quarter, so that we have $P(D1,D3)=9/16$. In a similar manner, we conclude that
$P(D1,D4)=P(D2,D3)=1/16$. This leaves a probability $P(\gamma)=1/4$ for the annihilation to occur.
We can also check our calculation by evaluating $P(\gamma)$ directly. Upon passing a beamsplitter each 
wave packet is reduced by a factor of $1/\sqrt{2}$, and the reflected one acquires an additional 
phase of $\pi/2$. Thus, the norm of the part of the wave function passing through $B$ and $C$ 
is $1/4$, which is also the probability for the pair to annihilate.
\newline
Once again a possible contradiction is avoided by noting that the results $P(D2,D4)=0$ and $P(D2,D4)=1/16$
refer to different statistical ensembles. As in Sect. VIII the compatibility test is trivial. 
The condition $P(D2,D4)=p_1+p_2+p_3+p_4=0$ frustrates any attempts to ascribe 
non-negative probabilities $p_i$, $i=1,2,3,4$ to the virtual paths in Eq.(\ref{h1}).
The uniquely quantum manner in which a new ensemble may be fabricated by introducing 
nothing more than a possibility of annihilation, remains the only "paradoxical" feature of the experiment. 
\section{Composite systems. The "quantum Cheshire cat"}
Our last example involves the so called "quantum Cheshire cat" model \cite{CAT1}, \cite{CAT2} 
still discussed in the literature \cite{CAT3}. The authors of \cite{CAT3} suggested that 
{\it "most of the works criticizing the effect did not introduce a proper framework in order to
analyze the issue of spatial separation of a quantum particle from one of its properties"},
so we take another look at the problem. 
Our more detailed 
analysis can be found in \cite{CAT4}, \cite{CAT5},  and here we will discuss a minimalist version of the model, 
which captures its essential properties.
\newline
Consider a composite, consisting of two two-level systems,  $S$ and $S'$, 
whose basis states are $|i\ra$ and $|i'\ra$, $i,i'=1,2$,  respectively. The full basis consists of four states, 
which we can choose to be the products $|i\ra|i'\ra$. We expand an arbitrary state of the composite as
$|\psi\ra =\alpha_1|1\ra|1'\ra+\alpha_2|1\ra|2'\ra+\alpha_3|2\ra|1'\ra+\alpha_4||2\ra|2'\ra$, 
thus representing it 
by a vector $\vec \alpha=(\alpha_1,\alpha_2,\alpha_3,\alpha_4$), $\sum_{j=1}^4|\alpha_j|^2=1$. We will need two commuting operators,
\begin{eqnarray}\label{r1}
\hat \Pi(1)=|1\ra|1'\ra \la 1'|\la1|+|1\ra|2'\ra \la 2'|\la1|\leftrightarrow \text {diag}(1,1,0,0)\q
\end{eqnarray} 
and 
\begin{eqnarray}\label{r2}
\hat \pi(2,2')=|2\ra|2'\ra \la 2'|\la2|\leftrightarrow \text {diag}(0,0,0,1).
\end{eqnarray}
(The authors of \cite{CAT1} used a slightly different operator, but $\hat \pi(2,2')$
is all we need for our analysis). 
\newline
Consider first the measurements with no post-selection made.
Suppose we make $N>>1$ measurements of $\hat \Pi(1)$ on a system prepared in the 
same state $|\psi\ra$. A result "1" will then be obtained with a probability $P(1)$
\begin{eqnarray}\label{r3}
P(1)=\la \psi|\hat \Pi|\psi\ra=|\alpha_1|^2+|\alpha_2|^2,
\end{eqnarray}
or in approximately $P(1)\times N$ trials. If instead we decide to measure $\hat \pi(2,2)$, 
the result "1" will occur with a probability 
\begin{eqnarray}\label{r4}
p(1)=\la \psi|\hat \pi(2,2)|\psi\ra=|\alpha_4|^2.
\end{eqnarray}
It is readily seen that if measuring $\hat \Pi(1)$ will yield "1" with certainty, $P(1)=1$,
measuring $\hat \pi(2,2)$ will only yield zeros, $p(1)=0$, 
\begin{eqnarray}\label{r4a}
P(1)=1 \to p(1)=0.
\end{eqnarray}
Obtaining a result "1" while measuring $\hat \pi(2,2')$ means that $S$ and $S'$ have been found
in the states $|2\ra$ and  $|2'\ra$, respectively. Obtaining a "1" while measuring $\hat \Pi(1)$ means that
$S$ has been found in $|1\ra$, regardless of where $S'$ might have been at the time.
It is not unreasonable, therefore, to interpret Eq.(\ref{r4a}) as a proof that
a system cannot "be found with certainty in two places (two orthogonal states) at the same time".
\newline
Next we ask what would change if the same measurements are to be made on a system later found (post-selected)
in some state $|\phi\ra$? It was already mentioned in Sect. V that the past and present of a quantum system 
have to be treated differently.  
In particular, passing through orthogonal states in the past does not, as such,  constitute exclusive alternatives.
Let the system make a transition between states $|\psi\ra$ and $|\phi\ra$, represented by coefficients
$\vec \alpha$ and $\vec \beta$, respectively. There are four virtual paths, shown in Fig. 11a,
\begin{eqnarray}\label{r5}
\{ 1\}\q \text{via}\q  |\phi\ra   \gets |1\ra|1'\ra  \gets |\psi\ra,\n
\{ 2\}\q \text{via}\q  |\phi\ra   \gets |1\ra|2'\ra  \gets |\psi\ra,\n
\{ 3\}\q \text{via}\q  |\phi\ra   \gets |2\ra|1'\ra  \gets |\psi\ra,\n
\{ 4\}\q \text{via}\q  |\phi\ra   \gets |2\ra|2'\ra  \gets |\psi\ra,
\end{eqnarray}
with the amplitudes $A(i)=\beta_i^*\alpha_i$ (we put $\HH=0$).
\begin{figure} 
	\centering
		\includegraphics[width=8cm,height=4cm]{{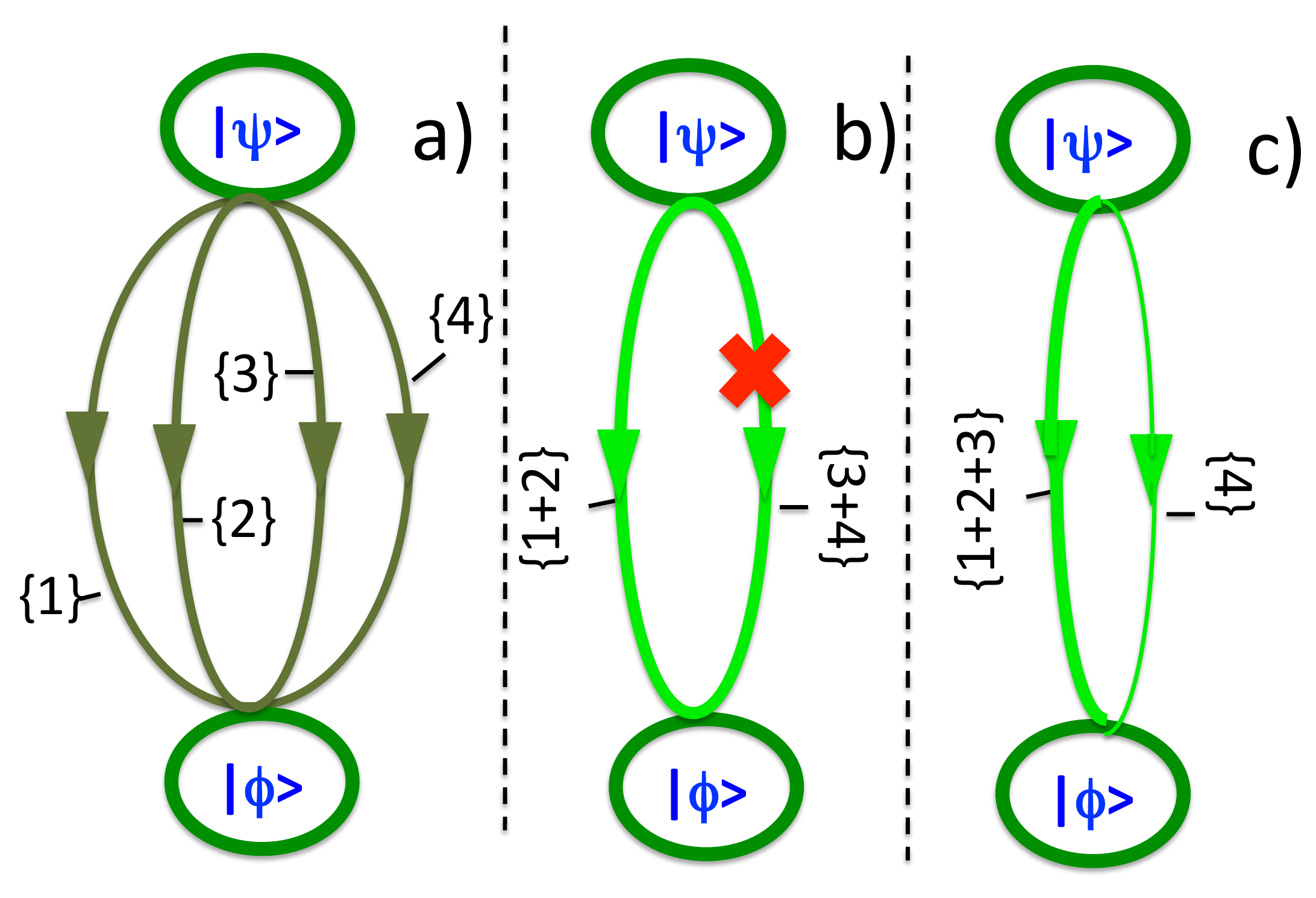}}
\caption{(Colour online) a) Four virtual paths in Eq.(\ref{r5}) connecting the initial and final sates 
of a composite system $S+S'$. b) Real paths produced by the measurement of the projector $\hat \Pi(1)$ 
in Eq.(\ref{r1}). c) Real paths produced by the measurement of the projector $\hat \pi(2,2')$ 
in Eq.(\ref{r2})}
\label{fig:12}
\end{figure}
\noindent
An intermediate measurement of  $\hat \Pi$ with two degenerate eigenvalues creates two real paths, $\{ 1\cup 2\}$ and $\{3\cup 4\}$ (see Fig. 11b), 
and a result "1" will occur with a probability
(we only keep the statistics if  the system is found in $|\phi\ra$ at the end) $P(1)=|A(1)+A(2)|^2
/[|A(1)+A(2)|^2+|A(3)+A(4)|^2]$.
Measuring instead $\hat \pi(2,2')$ creates paths $\{ 1\cup 2\cup 3\}$ and $\{4\}$ (see Fig. 11c), 
and a result "1" will occur with a probability
$p(1)=|A(4)|^2/[|A(1)+A(2)+A(3)|^2+|A(4)|^2$. Now choosing 
\begin{eqnarray}\label{r5a}
A(1)=A(2)=A(3)=-A(4)\ne 0
\end{eqnarray}
(suitable $|\psi\ra$ and $|\phi \ra$ can always be found 
\cite{{CAT4}}) we have 
\begin{eqnarray}\label{r4b}
P(1)=1 \q \text{and} \q p(1)
\ne 0.
\end{eqnarray}
Does this mean that, with post-selection, a particle can {\it "be found with certainty in two places at the same time"}?
As in all other examples considered above the answer is no. Compatibility test of Sect. III shows that it is not possible
to assign meaningful probabilities $p_i$, $i=1,2,3,4$ to the paths in Eqs.(\ref{r5}) , consistent with Eq.(\ref{r4a}). Indeed, the conditions
\begin{eqnarray}\label{r6}
p_1+p_2+p_3+p_4=1,\q\n
p_1+p_2=1,\q\q\q\q\q\n
p_4>0, \q\q\q\q\q\q\q
\end{eqnarray}
would imply $p_3<0$. With the results of two measurements incompatible, we can only make a trivial conclusion.
In one setup the final state can be reached via two routes, but one of them is not travelled. 
In a different setup the same state can be reached by two different routes, such that both are travelled 
with certain probabilities. Both setups are, in principle, possible.
\section{A note on "weak measurements"}
In the previous Sections we discussed the situations which may appear "paradoxical", but cease to be so 
once we realise that they refer to different statistical ensembles, produced in a peculiar quantum way, 
typically relying on non-local properties of quantum mechanics. The ensembles are not compatible in the sense
that each one requires a different experimental equipment, or different types of particles, and cannot be realised all at one time. 
Still, one may suspect that, since all experiments are performed on the same quantum system, different observable properties
must, in some sense, be "present" if the original system is taken in isolation. 
\newline
The suspicion is correct, and quantum mechanics provides a unique answer. A quantum system in isolation is characterised 
by probability amplitudes which can be ascribed to every plausible scenario. The knowledge of the amplitudes by no means implies
that a particular scenario {\it has been} realised, rather  they suggest what {\it would} happen if the 
corresponding
 equipment 
{\it were} installed in the lab and turned on. One should not be surprised by the fact that parts and combinations of these amplitudes 
can actually be measured in practice, e.g., by studying the response of a system to small perturbations, too weak to 
destroy coherence between interfering paths. Elementary perturbation theory routinely uses complex valued off-diagonal 
matrix elements of Hermitian operators \cite{Bohm}, and we may expect something similar to happen, should we decide to use meters which 
produce very little disturbance.
\newline
One way to minimise the perturbation is to make the initial state of a meter $G(f_k)$ in Eq.(\ref{b3}) to be  very broad in the coordinate space, e.g., 
by sending $\Delta f \to \infty$ in Eq.(\ref{ap1a}) of the  Appendix A. 
The large uncertainty in the initial pointer positions means that the measurement is highly inaccurate.
All $G$'s in Eq.(\ref{b3}), which are now nearly equal, amount to a common factor
in front of the sums, and the interference between all virtual paths is visibly restored. The problem is that all individual final values of $f_k$ are 
now almost equally probable, in accordance with the uncertainty principle \cite{FeynL}, 
which forbids telling apart interfering alternatives \cite{DSann}, \cite{DSmath}.  
The averages over many runs of the experiment are, however, well defined. 
In the absence of probabilities, these have to be expressed in terms of the amplitudes for the relevant virtual paths. 
\newline 
The rule, re-derived in the Appendix B, is quite simple. If the system is Sect. V passes from a state $|\psi_I\ra$ to some final state $|\phi\ra$, 
and at $t_4 >t_C >t_B> t_0$ we make inaccurate intermediate measurements of operators
\begin{eqnarray}\label{g0}
\B=\sum_{s_B=1}^{N_B}B_{s_B}\pb_{s_B},\q \text{and} \q \C=\sum_{s_C=1}^{N_C}C_{s_C}\pc_{s_C} \n
\end{eqnarray}
the mean pointer positions will be given by 
\begin{eqnarray}\label{g1}
\la f_B\ra =\text{Re}\left [\sum_{s_B=1}^{N_B}B_{s_B} \alpha (\phi,s_B|\psi_I)\right ]+O(1/\Delta),\n
\la f_C\ra =\text{Re}\left [\sum_{s_C=1}^{N_C}C_{s_C} \alpha (\phi,s_C|\psi_I)\right ]+O(1/\Delta),\n
\end{eqnarray}
where 
\begin{eqnarray}\label{g2}
\alpha (\phi,s_Z|\psi_I)= \frac{\la \phi|\U(t_4,t_Z)\hat \pi^Z_{s_Z}\U(t_z,t_0)|\psi_I\ra}{\la \phi|\U(t_4,t_0)|\psi_I\ra}, \q Z=B,C\q\q\n
\n
\end{eqnarray}
is the {\it relative amplitude} ($\sum_{s_Z=1}^{N_Z}\alpha (\phi,s_Z|\psi_I)=1$)  for reaching $|\phi\ra$ by passing first through the sub-space
$s_Z$. Note that two "weak" pointers are not disturbed by each other's presence,  and $\la f_C\ra$ is the same, as it would be 
if $\B$ were not measured at all. This situation is typical in perturbation theory, where the first-order effects are always additive.
\newline
The complex quantities in the square brackets in Eqs.(\ref{g1}) are nothing but linear combinations of quantum mechanical amplitudes \cite{CAT5}.
For example, after many trials, a "weak measurement" of a projector $\pb_{s_B}$ will let us evaluate the real part of the amplitude 
$\alpha (\phi,s_B|\psi_I)$ and, if $\pb_{s_B}=|b_s\ra \la b_s|$ projects onto a single state $|b_i\ra$, the real part of the amplitude 
for reaching  $|\phi\ra$ via $|\phi\ra \gets |b_s\ra \gets |\psi_I\ra$. Not that the imaginary parts of the $\alpha (\phi,s_Z|\psi_I)$ can be indirectly
measured in a slightly modified experiment \cite{Ah1}, but we are not concerned with them here. 
Rather, our purpose is to demonstrate that "weak measurements" 
fail to provide a deeper insight into the quantum "paradoxes" mentioned above.
\subsection {The three-box paradox of Sect. VIII.}
\noindent
Now weak measurements of $\B=\text{diag}(1,0,0)$ and $\C=\text{diag}(1,0,0)$
would yield the values
\begin{eqnarray}\label{g3}
\text{Re}[A(1,1,|\psi_I)/\sum_{j=1}^3A(1,j,|\psi_I)]=1,\q \text{and} \q
 \text{Re}[A(1,3,|\psi_I)/\sum_{j=1}^3A(1,j,|\psi_I)]=1.\q\q\q
\end{eqnarray}
But we already know this, for in tuning the system we had to ensure that $A(1,1,|\psi_I)=-A(1,2,|\psi_I)=A(1,3,|\psi_I)$,
or no "paradox" would arise.
Hence, no new information is gained. Neither does it mean that 
the system is {\it "...with certainty in one box, and with certainty in another box"} \cite{3Bb}.
The correct, if less intriguing,  statement would be " we checked that the real parts of the (relative) amplitudes for the two virtual pathways are both $1$".
\subsection{The nested interferometer of Sect. IX.}
\noindent
Here weak detectors placed at $A$ and $B$ will let us check previously known [cf. Eq.(\ref{e2})] relations
$\text{Re}[\alpha(1)]=\text{Re}[A(1)/\sum_{i=1}^3 A(i)]=\text{Re}[A(1)/A(3)]$  and $\text{Re}[\alpha(2)]=\text{Re}[A(2)/A(3)]=-\text{Re}[\alpha(1)]$.
Weak detectors placed only at $E$ and $F$, would both yield the real parts of the relative amplitude for the union of the paths $\{1\}$ and $\{2\}$, set to be zero, $\alpha(1\cup 2)=\alpha(1)+\alpha(2)=0$. The authors of \cite{Nest1} and \cite{Nest2} chose to conclude 
that, with no accurate intermediate measurements made, the photon never passed through $E$ and $F$, yet
"was" in $A$ and $B$, so its trajectory must be "discontinuous". Again, this is unwarranted. Their results only confirm the existence of the above relation between the amplitudes for passing via different arms, something the authors of \cite{Nest1} and \cite{Nest2} must have known when they were building their interferometer.
\subsection{ Hardy's paradox of Sect. XI.} 
\noindent
Here one can employ three commuting projectors $\hat \pi_i$,  $i=1,2,3$, associated with the three paths in Eqs.(\ref{h3}).  Weak measurements of these operators will yield relations between three relative amplitudes $\alpha(i)=A(i)/\sum_{i=j}^3 A(j)$, $i=1,2,3$, 
\begin{eqnarray}\label{g4}
\alpha(1)=-\alpha(2)=\alpha(3)=1,
\end{eqnarray}
already known from Eq.(\ref{h2}).
The authors of \cite{HARDY2} chose to  (i) associate with the quantities in Eqs.(\ref{g2}) {\it "the numbers of particle-antiparticle pairs"}, which pass through a particular route on the way to their respective detectors, and then discuss the meaning of the apparently negative number ($-1$) of such pairs passing through the outer arms of the interferometer.
It was also argued (ii) that the values in Eqs.(\ref{g2}) {\it "obey a simple, intuitive and, most importantly, 
self consistent logic"}. 
\newline
We note that (i) is unwarranted.  The relative probability amplitudes should not be interpreted as particle numbers.
One would not want to "explain" the two-slit diffraction experiment by saying 
that an infinite number of electrons arrive at a dark spot on the screen through one slit, and minus the same number of electrons reach it via the other slit. Furthermore, the statement (ii) is trivial. Relative amplitudes abide by the simple logic obeyed by all probability amplitudes \cite{FeynL}, plus the additional condition that they should add up to unity. 
\newline
\subsection{The quantum Cheshire cat of Sect. XII.} 
\noindent
In this case, the relative amplitudes, probed by 
weak measurements of the operators $\hat \Pi(1)$ in Eq.(\ref{r1}) and $\hat \pi(2,2')$ in Eq.(\ref{r2}), are 
given by
\begin{eqnarray}\label{g5}
\alpha(1\cup 2)=1, \q \alpha(4)=A(4)/[A(1)+A(2)+A(3)]\ne 0, 
\end{eqnarray}
as one should expect from Eq.(\ref{r5a}). This is all an analysis in terms of "weak measurements" can add to our discussion. The authors of \cite{CAT1} chose instead to speak about {\it "separation of a particle from its properties"}, 
and we leave it to the reader to decide whether Eqs.(\ref{g5}) warrant such an interpretation.
\newline
Finally, we emphasise the circular nature of the argument in all of the above examples. A system is tuned so as to ensure cancellation between certain amplitudes. Then weak measurements are invoked to demonstrate that the amplitudes do indeed have the 
desired properties. Calling amplitudes "weak values" does no particular harm, although adds little to the discussion. 
Expecting them to be more than just amplitudes, is likely to lead to confusion, and ought to be avoided.
\section{Conclusions and discussion}
Our main purpose  was to look for a straightforward description of a quantum system subjected to several intermediate measurements.  and here we present our findings.
Accurate measurements in a finite-dimensional Hilbert space produce sequences of discrete outcomes, given by the eigenvalues of the measured operators. Thus, the result is a classical statistical ensemble, 
specified by the initial and final states of the system,  the "real" pathways which connect these states, and the probabilities
with which the pathways are taken. Different sets of measurements "fabricate" different ensembles, which may have little in common, except for the quantum system used in their production. 
\newline
The uniquely quantum nature of the problem is evident from the manner in which an ensemble is produced, 
since the observed real pathways can only be constructed after referring first to the virtual paths  of the unobserved system. As a rule, there are many virtual paths, and the relevant ones are selected by the measurements which are to be made. An accurate intermediate measurement then destroys interference between virtual paths to a degree which depends 
on the degeneracy of the eigenvalues of the measured operators.
\newline
For a given number of intermediate measurements, the maximum number of real pathways, and the most detailed (fine-grained) ensemble, are produced when all eigenvalues are distinct. A less detailed (coarse-grained) ensemble with fewer possible outcomes results from measuring operators 
with degenerate eigenvalues. 
 Typically, a less detailed ensemble cannot be produced from the most detailed one by classical coarse-graining, i.e., by simply adding the probabilities wherever the eigenvalues turn out to be identical. 
A simple compatibility test can often be used to demonstrate that, in order to yield the correct coarse-grained probabilities,
some of the fine-grained probabilities would have to be negative, and a suitable fine-grained classical ensemble simply does not exist.
This can be seen as a generalisation of the well known principle, which states that the results of a single quantum measurement 
cannot be pre-determined, and are produced in the course of the measurement.
\newline
Coarse graining must, therefore, occur via uniting virtual pathways, and adding their amplitudes, rather than probabilities.
Since amplitudes may have opposite signs, the union of two paths may have a zero probability, even though both paths 
would be travelled if interference between them were destroyed. Such is, for example, the case of a photon arriving,
after one slit has been blocked,  
at the (formerly) dark spot of the interference pattern.
This uniquely quantum feature allows one to produce essentially different statistical 
ensembles from the same quantum system, and makes the results of different sets of measurements incompatible, 
as discussed in Sect. III.
\newline
Another important distinction between fabrication of an ensemble by purely classical and by quantum means, 
is the use of non-locality inherent in quantum mechanics. It can be possible to make a classical stochastic system which would reproduce the probabilities of  quantum measurements. It can be possible to alter it, in order 
for it to  mimic a different choice of quantum measurements. However, where an alteration may require several changes made at different locations of the classical setup, fewer modifications could suffice in the quantum case.
For example, what would require flipping the switch  $SW$ and resetting the connector $Br1$  in Fig. 9, can be achieved by simply flipping the switch $SW$ in Fig. 8. 
\newline
The layout of the real pathways does not aways coincide with the physical layout of the system.
Thus, in the absence of intermediate measurements, different arms of interferometer in Fig. 7a are, in fact, virtual paths, and the only real pathway via which a photon can reach the detector is their superposition shown in Fig. 7b.
Similarly, in the Hardy's example, the system in Fig. 10a consists of four distinct wave guides, while the virtual paths of interest 
are described in Eqs. (\ref{h1}). 
With the path $\{4\}$ out of the game, the remaining three interfere to 
produce a single real pathway $\{1\cup 2\cup 3\}$ in Fig. 10c, thus enabling detection by both $D2$ and $D4$.
\newline 
There is one aspect in which a quantum-made ensemble resembles a purely classical one.
For a chosen set of "ensemble forming" measurements, there exist other "non-perturbing" measurements which, if added, will not alter the original probabilities, create new real pathways, or change the system's possible destinations. Unique values, obtained in such measurements, play the role of the pre-existing values of classical physics.  With non-perturbing measurements, and only with them,  the situation is indeed classical, 
and observation without disturbing is possible. For example, in the Hardy's case, 
detectors designed to distinguish between the path $\{4\}$ leading to annihilation, and the paths $\{1\}$, $\{2\}$ and $\{3\}$, 
but not between $\{1\}$, $\{2\}$ and $\{3\}$ individually, could be installed at several locations along the wave guides.
All detectors would click without fail, whenever the pair of particles  is detected by  $D2$ and $D4$, which would still happen 
with the probability $1/16$. 
It is in this sense that the pathway uniting the three virtual paths in Eq.(\ref{h3})
 can be considered real.
\newline
A yet more detailed analysis, desirable as though may seem, encounter a serious difficulty.
The wave guides, which to us are three different objects, appear but  a single conduit  for the particle-antiparticle pair. Is it  possible to peek inside this "real path", 
and see what {\it actually} happens in different arms of the setup shown in Fig. 10a?
The uncertainty principle says no \cite{DSmath}, 
and in trying to answer the "which way?" question we meet with a familiar dilemma \cite{FeynL}.
Installing one or more accurate detectors 
would tell which way the pair has travelled, but would also change the probability of  detection by $D2$ and $D4$ from
$1/16$ to $3/16$, thus leaving us with a different ensemble. 
Making the detectors as delicate as possible helps to avoid the perturbation,
but only reveals relations between probability amplitudes, in principle already known to us.
We ask "what can be said about individual virtual paths?" and receive,
in an operational way, the textbook answer "probability amplitudes".
\newline
Finally, the recipe for evaluating frequencies of observed events by finding first the amplitudes of all 
relevant virtual scenarios, adding them up, as appropriate, and then taking the absolute squares 
is by no means new. It can be found in the first chapter of Feynman's undergraduate text \cite{FeynL}.
Above we only outlined a particular way to define the scenarios, which may occur in consecutive measurements, and the rules for adding their amplitudes. 
Our conclusions may seem, therefore, to point towards the
{\it "Shut up an calculate!"} attitude, famously articulated by David Mermin \cite{Merm2}. 
Could we have done better, for example, by providing a yet deeper insight into why the measurements outcomes 
are random, and why is it necessary to deal with complex valued probability amplitudes, before the probabilities of the outcomes can even be evaluated? Probably not.
We conclude with Feynman's advice to his undergraduate audience \cite{FeynL}:
{\it So at the present time we must limit ourselves to computing probabilities. We say "at the present time", but we suspect very strongly that it is something to be with us forever...- that this is the way nature really is.}
\section{Appendix A.  Accurate ("Strong") impulsive von Neumann measurements}
An impulsive von Neumann measurement \cite{Real3} of an operator $\B=\sum_i|b_i\ra B^\la b_i|$ at $t=t_1$ entangles the state of the measured system, $|\psi\ra=\sum_i^N \la b_i|\psi\ra |b_i\ra $, with the position of a pointer, $|f\ra$, 
by means of an interaction Hamiltonian $\HH_{int}= -ig \partial_f\B \delta(t-t_1)$. 
Initially, the state of the composite system+meter,  $|\Phi\ra$, is the product of $|\psi\ra$, and the pointer's state $|G\ra$. With the constant $g$ put to unity, immediately after the interaction $|\Phi\ra$ is transformed into $|\Phi'\ra$,
$\la f|\Phi'\ra=\sum_iG(f-B_i) \la b_i|\psi\ra |b_i\ra$, where $G(f)=\la f|G\ra$. One extracts information about the value of $\B$, by accurately determining 
the pointer's final  position. 
\newline Next consider $K$ such measurements, performed at $0 <t_1<t_2...<t_K<T$ on a system with an evolution operator $\U$.
The measured operators $\B(k)=\sum_{i_k=1}^N |b_{i_k}\ra B^{i_k}(k) \la b_{i_k}\ra$ need not commute and, for simplicity, we assume their eigenvalues to be non-degenerate. In addition, at $t=T$ the system is found (post-selected) in a state $|\phi\ra$. Performing the measurements one after another, we find the amplitude to have the pointer readings $f_k$, $k=1,K$, 
\begin{eqnarray}\label{ap1}
A(\phi,f_K,...,f_1)=\sum_{i_1...i_K}\prod_{k=1}^KG(f_k-B^{i_k}(k))
A(\phi,i_K,...,i_1),\q
\q\n
A(\phi,i_K,...,i_1)\equiv \la \phi | \U(T,t_k)|b_{i_K}\ra\la b_{i_K}|...|b_{i_1}\ra
 \la b_{i_1}|\U(t_1,0)|\psi\ra,\q
\end{eqnarray} 
with the corresponding probability given by $P(\phi,f_K,...,f_1)=|A(\phi,f_K,...,f_1)|^2$.
\newline
Let all $G(f_k)$ be identical Gaussians of a width $\Delta f$,
\begin{eqnarray}\label{ap1a}
G(f_k)=2^{1/4}\pi^{-1/4}\Delta f^{-1/2}\exp(-f_k^2/\Delta f^2). 
\end{eqnarray} 
For an accurate (strong) measurement we choose $\Delta f \to 0$, which gives 
\begin{eqnarray}\label{ap1aa}
G(f_k-B^i(k))G(f_k-B^j(k)\to
\delta(f_k-B^i(k))\delta_{ij}, 
\end{eqnarray}
and, with the help of Eq. (\ref{ap1}), for the probability to find pointers readings $f_1,f_2,...,f_K$ we obtain
\begin{eqnarray}\label{ap2}
P(\phi,f_K,...,f_1)=
\frac
{\sum_{i_1...i_K}\prod_{k=1}^K \delta(f_k-B^{i_k}(k))|A(\phi,i_K,...,i_1)|^2}{\sum_{i_1...i_K}
  |A(\phi,i_K,...,i_1)|^2}.\q\q\n
\end{eqnarray} 
Thus, the average position of the $n$-th pointer is given by,
\begin{eqnarray}\label{ap3}
\la f_n\ra \equiv \prod_{k=1}^K df_k f_n P(\phi,f_K,...,f_1)=
\sum_{i_1...i_K=1}^NB^{i_n}|A(\phi,i_K,...,i_1)|^2/\sum_{i_1...i_K=1}^N|A(\phi,i_K,...,i_1)|^2,\q\q\n
\end{eqnarray}
and for a correlator $\la \prod_{k=1}^K f_k\ra$ we find
\begin{eqnarray}\label{ap4}
\la \prod_{k=1}^K f_k\ra \equiv
\frac{\sum_{i_1...i_K=1}^N\prod_{k=1}^K B^{i_k}|A(\phi,i_K,...,i_1)|^2}{\sum_{i_1...i_K}
  |A(\phi,i_K,...,i_1)|^2},\q\q\n
\end{eqnarray}
Clearly, there are $N^K$ real pathways, $\phi \gets i_K \gets ... \gets i_1\gets \psi$, 
which the measured system travels with the probabilities $|A(\phi,i_K,...,i_1)|^2$. The value of $\B(k)$ on a pathway 
passing through $i_k$ is its eigenvalue $B^{i_k}(k)$. 
\newline
The approach is easily extended to measurements of operators with  $N_k$ degenerate eigenvalues, $\B(k)=\sum_{s_k=1}^{N_k }B_{s_k}(k) \hat \pi_{s_k}$, where $ \hat \pi_{s_k}$ is the projector on the sub-space spanned by the eigenstates, corresponding to 
the degenerate eigenvalue $B_{s_k}(k)$.  In this case, in Eqs. (\ref{ap1}) $B^{i_k}$ should be replaced by
$B_{s_k}(k)$, summation over $i_k$ by a sum over $s_k$, and one-dimensional projectors $|b_{i_k}\ra \la b_{i_k}|$ by $\hat \pi_{s_k}$.
A useful example is the measurement of a series of one-dimensional projectors, $\B(k)=|b_{m_k}\ra \la b_{m_k}|$. Now each $\B(k)$ 
has a non-degenerate eigenvalue $B_1=1$, and $(N-1)$-degenerate eigenvalue $B_2=0$, so that $S_k=2$. There are $2^K$ real pathways, passing at $t=t_k$ either through the state $|b_{m_k}\ra$, or through the sub-space orthogonal to it. It is easy to see that that 
the average in Eq.(\ref{ap4}) now coincides with the probability to travel the path passing through $|b_{m_k}\ra$ at every $t_k$,
\begin{eqnarray}\label{ap4a}
\la \prod_{k=1}^K f_k\ra=\frac{|A(\phi,m_K,...,m_1)|^2}{\sum_{i_1...i_K}
  |A(\phi,i_K,...,i_1)|^2}.
\end{eqnarray}  
 \b{ Finally, in practice it may be more convenient to couple the pointer via its position $f$, rather than through its momentum $\lambda$, and measure its final velocity instead of its final position. With the interaction now reading $\HH_{int}= f\B \delta(t-t_1)$, a change to the 
momentum representation (with respect to the pointer) yields $\HH_{int}= i\partial_\lambda \B \delta(t-t_1)$. 
After the interaction with the meter we have $\la \lambda |\Phi'\ra=\sum_iG(\lambda+B_i) \la b_i|\psi\ra |b_i\ra$, where 
$G(\lambda)=\la \lambda |G\ra$. A strong accurate measurement now requires a pointer with a well defined initial momentum
(a narrow $G(\lambda)$), and the only other difference is the sign of the shift in the momentum space evident in $G(\lambda+B_i)$.} 
\section{Appendix B. Inaccurate ("weak") impulsive von Neumann measurements}
\b{A natural complement to the Appendix A is a brief discussion of the results obtained in the limit where the measurements are made deliberately inaccurate, in order to perturb the observed system as little as possible. 
In \cite{Georg} the authors, who studied a similar problem, chose to reduce the couplings between the system and the meters involved. Here we would do it slightly differently, namely by sending $\Delta f$ in Eq.(\ref{ap1a}) to infinity, and making the initial position of a meter indeterminate.}
It is convenient to rewrite Eq.(\ref{ap1}) as 
\begin{eqnarray}\label{bp1}
A(\phi,f_K,...,f_1)=\int_{-\infty}^\infty df_1'...df_K'
\prod_{k=1}^KG(f_k-f_k')\eta(f_1',...f_K')\q\q\n
\eta(f_1',...f_K')\equiv \sum_{i_1...i_K=1}^N
\prod_{k=1}^K \delta(f_k'-B^{i_k}(k))A(\phi,i_K,...,i_1).\q
\end{eqnarray} 
It is readily seen that
\begin{eqnarray}\label{bp2}
\int df_1...df_K \eta(f_1,..f_K)=\la \phi | \U(T,0)|\psi\ra\equiv A(\phi). 
\end{eqnarray} 
Since integration over $f'_k$ is restricted to the finite range, containing the eigenvalues of $\B(k)$, 
and $G(f)$ is now broad, we can write 
\begin{eqnarray}\label{bp3x}
G(f_k-f_k') \approx G(f_k)-G'(f_k)f_k',
\end{eqnarray}
where $G'(f) \equiv dG(f)/df$ is a small parameter.
Individual pointer positions can now lie almost everywhere, and are of no particular interest \cite{DSmath}.
Whatever information about $\B(k)$ such a measurement may produce, should be obtained from 
the average pointer positions, evaluated over many runs of the experiment. Let us see what kind of information 
can be gained in this way.
For $\la f_k\ra$, to the leading order in $1/\Delta f$, we have
\begin{eqnarray}\label{bp3}
\la f_k\ra = |A(\phi)|^2\text{Re} \left [\frac{\sum_{i_k=1}^N B^{i_k}(k) A(\phi,i_k)}{\sum_{i_k=1}^N A(\phi,i_k)}\right ],\n
A(\phi,i_k) \equiv \sum_{i_1...i_{k-1}i_{k+1}...i_K}A(\phi,i_K,...,i_1)=
\la \phi |\U(T,t_k)|b_{i_k}\ra\la b_{i_k}|\U(t_k,0)|\psi\ra, 
\end{eqnarray}
where we have used 
\begin{eqnarray}\label{bp4}
\int f G^2(f)=0,\q\q\q\q\n 
\int f G'(f)G(f)df=\int f (G(f)^2)'df/2= 
-1/2,\n
\int df_1...df_K f_k \eta(f_1...f_K)=\sum_{i_k=1}^N B^{i_k}(k) A(\phi,i_k),
\end{eqnarray}
and Eq.(\ref{bp2}). Equation (\ref{bp4}) is valid regardless of whether the eigenvalues
$B^{i_k}(k)$ are degenerate or not.
 The complex valued quantity in rectangular brackets in Eq.(\ref{bp3}) is often called "the weak value of an operator 
$\B(k)$", which can also be written in an equivalent form (see, e.g., \cite{Nest6})
\begin{eqnarray}\label{bp5}
\la \B(k)\ra_w = \frac{\la \phi | \U(T,t_k) \B(k) \U(t_k,0)|\psi\ra}{\la \phi | \U(T,0)|\psi\ra}
\equiv  \sum_{i_k=1}^N B^{i_k}(k) \alpha(\phi,i_k).
\end{eqnarray}
where  $\alpha(\phi,i_k)= A(\phi,i_k)/\sum_{i_k=1}^N A(\phi,i_k)$ is the {\it relative} amplitude for reaching $|\phi\ra$ via
$|\phi\ra  \gets |b_{i_k}\ra \gets |\psi\ra$. The weak value (\ref{bp5}) is, therefore, a sum of such relative amplitudes, weighted by the eigenvalues of $\B$.
The weak value of a projector $\hat \pi_{\mathcal{M}}$  onto a sub-space, $\mathcal{M}$, spanned by orthogonal states $|b_{m_k}\ra$, 
$m_k=1,2,...M_k$,
\begin{eqnarray}\label{bp5a}
\hat \pi_{\mathcal{M}}=\sum_{m_k=1}^{M_k}|b_{m_k}\ra \la b_{m_k}|\
\end{eqnarray}
 is just the sum of the  relative amplitudes for all paths $|\phi\ra  \gets |b_{m_k}\ra \gets |\psi\ra$, 
 \begin{eqnarray}\label{bp5b}
\la \hat \pi_{\mathcal{M}}\ra_w=\sum_{m_k=1}^{M_k} \alpha(\phi,m_k).
\end{eqnarray}
Calculation of correlations between positions of several weak pointers is more involved.
Using equations (\ref{bp4}), we have
\begin{eqnarray}\label{bp5bb}
\la \prod_{k=1}^Kf_k\ra\approx 2^{-2}\sum_{s_1,...,s_K=0,1}
\int \prod_{k=1}^K f_k^{s_k}\eta(f_1,...,f_K)df_1...df_k
\int \prod_{k=1}^K f_k^{|s_k-1|}\eta^*(f'_1,...,f'_K)df'_1...df'_k.\q
\end{eqnarray}
For just two weak meters, $K=2$, Eq.(\ref{bp5bb}) yields
\begin{eqnarray}\label{bp6}
\la f_1f_2\ra\approx 2^{-1}|A(\phi)|^2
\left \{ \text{Re}[\la \B(1)\ra^*_w\la \B(2)\ra_w]
+\text{Re}[\la \B(1)B(2)\ra_w] \right \}. 
\end{eqnarray} 
where $\la \B(1)\ra_w$ and $\la \B(2)\ra_w$ are given by Eqs.(\ref{bp5}).
The second term in Eq.(\ref{bp6}),  $\la \B(1)B(2)\ra_w$,  is the sequential weak value of  Georgiev  and Cohen \cite{Georg}
\begin{eqnarray}\label{bp7}
\la \B(1)B(2)\ra_w \equiv \frac{\sum_{i_1,i_2=1}^N B^{i_1}(1)B^{i_2}(2) A(\phi,i_1,i_2)}{\sum_{i_1,i_2=1}^N A(\phi,i_1,i_2)}
=\frac{\la \phi | \U(T,t_2)\B(1) \U(t_2,t_1)\B(1) \U(t_1,0)|\psi\ra}{\la \phi | \U(T,0)|\psi\ra}.
\end{eqnarray}
where $A(\phi,i_1,i_2) \equiv \la \phi | \U(T,t_2)|b_{i_2}\ra \la b_{i_2}|  \U(t_2,t_1)|b_{i_1}\ra$ $ \la b_{i_1}|\U(t_1,0)|\psi\ra$
is the amplitude to reach $|\phi\ra$ via $|\phi\ra \gets |b_{i_2}\ra \gets |b_{i_1}\ra \gets |\psi\ra$. With both $\B(1)$ and $\B(2)$ chosen to be the projectors onto $|b_{m_1}\ra$ and $|b_{m_2}\ra$, respectively, the sequential weak value (\ref{bp7}) reduces to the amplitude
$A(\phi,m_1,m_2)$, as was argued in \cite{Georg}. Note, however, that it is not directly proportional to the measured $\la f_1f_2\ra$, due to the presence of the extra term in the r.h.s. of Eq.(\ref{bp6}). 
\newline 
We finish  our discussion of weak measurements by noting the well known fact \cite{Ah1} that also imaginary parts of the weak values or, indeed, any real valued combination of their real and imaginary parts \cite{DSann}, can be measured 
by preparing and  reading the pointer(s) differently. For example, we can determine the average values of the pointer momenta, $\hat{\lambda}_k=-i\partial_{f_k}$, instead of their mean positions. Using 
\begin{eqnarray}\label{bp8}
\int G(f)G'(f)df =\int (G^2(f))'df/2=0,\n 
\int G(f)G''(f)df =-\int G'(f)^2 df=-1/\Delta f^2 
\end{eqnarray}
we obtain equations similar to  Eqs. (\ref{bp3}) and (\ref{bp6}), but for the imaginary parts of the quantities involved, 
\begin{eqnarray}\label{bp9}
\la \lambda_k\ra = 2(\Delta f)^{-2}|A(\phi)|^2\text{Im}\left [ \la \B(k)\ra_w\right ], 
\end{eqnarray}
and 
\begin{eqnarray}\label{bp10}
\la \lambda_1\lambda_2\ra\approx 4(\Delta f)^{-4}|A(\phi)|^2
 \left \{ \text{Im}[\la \B(1)\ra^*_w\la \B(2)\ra_w]
+\text{Im}[\la \B(1)B(2)\ra_w] \right \}. 
\end{eqnarray}
\b{ As in the case of strong measurements (see Appendix A), we could also use a pointer, coupled
to the system via  $\HH_{int}= i\partial_\lambda \B \delta(t-t_1)$. Then, to make the measurement inaccurate (weak), we would choose its initial 
state to have a large uncertainty in the momentum space (a broad $G(\lambda)$). Now evaluating its mean momentum $\la \lambda \ra$ will allow one to determine the real part of $\la \B(k)\ra_w$ in Eq.(\ref{bp5}), while its mean position, $\la f \ra$, will contain 
information about $\text{Im}[\la \B(k)\ra_w]$.}
\newline
The reader who finds the content of this Appendix but a tedious exercise in perturbation theory would be right.
Since weakly perturbing meters do not create new real pathways, there are no probabilities, and  their mean readings are expressed in terms of the corresponding probability amplitudes and their weighted combinations, otherwise
known as "weak values". 

 \section {Acknowledgements}  Support by the Basque Government 
 (Grant No. IT986-16), as well as by 
MINECO and the European Regional Development Fund FEDER, through the grant
FIS2015-67161-P (MINECO/FEDER) 
is gratefully acknowledged.

\end{document}